\documentclass[useAMS,usenatbib]{mn2e}
\usepackage{graphicx}

\title[Globular Clusters: DNA of galaxies ?]{Globular Clusters: DNA of 
Early-Type galaxies?}
\author[Forte et al.]{Juan C. Forte$^{1,2}$\thanks{E-mail: forte@fcaglp.unlp.edu.ar}, E. Irene Vega$^{2,3}$, Favio R. Faifer$^{2,3,4}$, Anal\'ia V. Smith Castelli$^{2,3,4}$, 
\newauthor 
Carlos Escudero$^{2,3,4}$, N\'elida M. Gonz\'alez$^{3,4}$,
 Leandro Sesto$^{2,3,4}$\\
$^1${Planetario ``Galileo Galilei'', Secretar\'ia de Cultura, Ciudad
  Aut\'onoma de Buenos Aires, Argentina}\\
$^2${Consejo Nacional de Investigaciones Cient\'ificas y T\'ecnicas,
  Av. Rivadavia 1917, C1033AAJ, Ciudad Aut\'onoma de Buenos Aires,
  Argentina}\\
$^3${Facultad de Ciencias Astron\'omicas y Geof\'isicas, Universidad
  Nacional de La Plata, Paseo del Bosque, B1900FWA, La Plata,
  Argentina}\\
$^4${Instituto de Astrof\'isica de La Plata (CCT-La Plata, CONICET-UNLP),
  Paseo del Bosque, B1900FWA, La Plata, Argentina}\\
}

\begin{document}

\date{Accepted ;  Received ; in original form }
\pagerange{\pageref{firstpage}--\pageref{lastpage}} \pubyear{}
\maketitle

\label{firstpage}

\begin{abstract}

This paper explores if the mean properties of Early-Type Galaxies (ETG) can 
be reconstructed from ``genetic'' information stored in their GCs (i.e., in 
their chemical abundances, spatial distributions and ages). This approach 
implies that the formation of each globular occurs in very massive stellar 
environments, as suggested by some models that aim at explaining the presence 
of multi-populations in these systems.  The assumption that the relative 
number of globular clusters to diffuse stellar mass depends exponentially on 
chemical abundance, [Z/H], and the presence of two dominant GC sub-populations 
({\it blue} and {\it red}), allows the mapping of low metallicity halos and of 
higher metallicity (and more heterogeneous) bulges. In particular, the masses 
of the low-metallicity halos seem to scale up with dark matter mass through a 
constant. We also find a dependence of the globular cluster formation 
efficiency with the mean projected stellar mass density of the galaxies within 
their effective radii. The analysis is based on a selected sub-sample of 
galaxies observed within the ACS Virgo Cluster Survey of the {\it Hubble Space 
Telescope}. These systems were grouped, according to their absolute 
magnitudes, in order to define composite {\it fiducial} galaxies and look for a 
quantitative connection with their (also composite) globular clusters systems. 
The results strengthen the idea that globular clusters are good quantitative 
tracers of both baryonic and dark matter in ETGs.

\end{abstract}
\section{Introduction}

\label{INTRO}
The characteristics of living creatures, as well known, can be linked to 
their DNA properties. In a similar fashion we may ask if, given the features 
of a Globular Cluster System (GCS) (defined in terms of spatial distribution, 
chemical composition and ages), they can yield ``genetic'' information about the 
field stellar populations of the Early-Type Galaxies (ETGs) they belong to. 
This question can be extended to late-type galaxies regarding the formation of 
their stellar halos and bulges.

A number of papers in the literature have pointed out differences between 
field stars and globular clusters (GCs) rather than  similarities  
\citep[e.g.][]{KISS09}. Although such differences should take into account 
that GCS properties come in {\it number weighted} form, contrasting with those 
of galaxy  properties, that we read in a {\it luminosity weighted} way, the 
generalized conclusion seems to indicate that GCs and field stars have 
followed different formation histories. 
 
In the meantime, the discovery of multi-stellar populations in GCs suggests 
processes where the GC formation was {\it ``a for more titanic event than ever 
imagined before''} \citep[see][]{REN13}. These scenarios involve large amounts 
of field stars per GC that would share some characteristics of the clusters 
\citep{DER08,CARR10}.

Despite some difficulties (e.g. the required high star formation efficiencies; 
see \citealt{BAS13}), these models seem consistent with the idea that GCs 
could be  tracers of much more massive diffuse stellar populations. This kind 
of approach was explored  in \citet[][hereafter, FFG05]{FFG05}, 
\citet[][FFG07]{FFG07}, \citet[][FVF09]{FVF09} and \citet[][FVF12]{FVF12}.

 In a recent paper, \citet[][HHA13 in what follows]{HAR13}, present an 
extensive compilation of several structural parameters that characterize GCS 
and their parent galaxies. These authors also discuss the implications of a 
number of correlations that arise considering GCs as a whole, i.e. not 
including the effects of GCs sub-populations. One of their conclusions 
indicates that the total number of GCs is not a good estimator of the total 
stellar mass of a galaxy. In fact, a linear {\it log-log} fit between these 
quantities shows deviations both at the low and high galaxy mass regimes.

They also show that, in a GC population so defined, there is a clear trend 
between the mass fraction locked in GCs,  $S_m$,  and the total stellar 
galaxy mass that adopts a {\it U-shaped} form, already noted by \citet{PEN08}, 
in terms of the {\it GC specific frequency}  $S_n$ (number of GCs per 
V-luminosity in $M_V=-15$ units). HHA13 interpret that this shape is the 
consequence of a decrease  of the star formation efficiency, both at the low 
and high mass regimes produced by mass loss and by the energetic events 
connected with the nuclei of galaxies, respectively.
 
In this paper we attempt to elaborate further on those results by introducing 
the effect of GC bi-modality, i.e., the presence of {\it blue} and {\it red} 
GCs. In particular, we aim at exploring the usefulness of these distinct 
GC populations to trace large scale features of galaxies as, total stellar 
mass, the relative contribution of halos and bulges to that mass, chemical 
abundances, stellar mass to luminosity ratios and, indirectly, dark matter 
content.  

A preliminary attempt to link galaxies with their GCS in terms of the 
blue and red populations, was presented in FVF09 for galaxies 
observed within the ACS Virgo Cluster Survey of the {\it Hubble Space 
Telescope} \citep{COT04}. In this work, we revise that approach taking 
advantage of later developments, namely, the accessibility to the original 
photometry of GCs in the ACS Virgo galaxies \citep{JOR09}, the multicolour 
photometry of these galaxies \citep{CHE10}, and the C-griz' 
colour-metallicity grid from \citet[][hereafter, F13]{FOR13}.
  
As in previous works, we stand on the idea that GC formation is not a singular 
event but, rather, is associated with large volumes of star formation that 
eventually lead to the origin of distinct populations: a low metallicity
and spatially extended stellar halo, associated with the blue globulars, 
and a more concentrated and chemically more heterogeneous stellar population, 
connected with the red GCs. The presence of extended low-metallicity
halos has been noticed by different means, for example, in NGC\,3379 
\citep{HAR07} and NGC\,4472 \citep{MIHHA13}.
  
This work assumes that the GCs colour bi-modality is the consequence of the 
genuine bimodal nature of their chemical composition distributions 
\citep[see][]{BRO12}, i.e., not an artifact of some eventually non-linear
integrated colour-metallicity relations. On the other side, and as noted in 
F13, this last kind of relations does not reject necessarily the possibility 
of a genuine chemical bi-modality.

The paper is organized as follows. The definition of {\it fiducial} galaxies 
is given in Section\,\ref{FIDUCIALS}. The analysis of the GCs  
colour distributions, as well as their modelling in terms of chemical 
abundance, 
are described in Section\,\ref{Color}. The link between GCs and the diffuse 
stellar populations of the galaxies is presented in Section\,\ref{FIELD}, while 
the photometric scales for globular clusters and galaxies are discussed in 
Section\,\ref{ZEROPOINTS}. The model results based on galaxy colours and 
integrated brightness are discussed in Section\,\ref{COLOURS}. The analysis of 
the shape of the relation between total number of GCs and galaxy masses is 
presented in Section\,\ref{BREAK}. The connection between the 
S\'ersic parameter and the projected stellar surface mass density 
for the fiducial galaxies is presented in Section\,\ref{Sersic}. The results 
of modelling the low-metallicity stellar halo, bulge-like component, and total 
stellar masses, as well as  the GC {\it formation efficiency} 
are presented in Section\,\ref{HALOBULGE}. The relation between the dark 
matter content of each galaxy and their low-metallicity stellar haloes is 
discussed in Section\,\ref{LOWMETDARK}. The case of the Fornax spheroidal 
galaxy is described in Section\,\ref{Fornax}, and the final conclusions are 
given in Section\,\ref{CONCLU}.


\section{The definition of fiducial galaxies and their GCS}
\label{FIDUCIALS}

One of the problems in looking for a GC-field stars connection is the noise 
inherent to the counting statistics of GCs, an effect that becomes more severe 
as the galaxy is fainter and the number of associated GCs is lower. In order 
to decrease this effect we ``merge'' both the stellar and GCs populations 
of Virgo ACS galaxies creating {\it fiducial} galaxies. For this purpose, we 
compose the $g$ and $z$ brightness (through the observed flux in each band) in 
order to derive the integrated magnitudes and $(g-z)$ colours of each fiducial 
galaxy.

This approach would work only if there is a pattern, common to all galaxies, 
linking the features of field stars and their GCS. The idea also assumes that 
the bulk of field stars and GCs are mostly coeval or spread over a limited 
range of ages. This seems to be true, at least for galaxies of the 
``red sequence'' (or see \citealt{NOR08}, for the  particular case of 
NGC\,3923).

The fiducial systems contain a given number of galaxies, ordered by decreasing 
brightness in the $g$ band, and were defined trying to fulfill two conditions. 
On one side, they should have a sufficiently large number of GCs to allow a proper fit of their composite colour histograms and, on the other, a significant separation in 
terms of the mean absolute magnitudes of consecutive fiducial galaxies (at 
least $\approx0.5$ mag). 

In this work we adopt the ugriz' galaxy photometry extensively discussed by 
\citet{CHE10} although we restrict our analysis only to the $(g-z)$ colour 
index. In the case of the GCs we use the  photometric $g$ and $(g-z)$ data 
presented by \citet{JOR09}, after taking into account completeness and field 
contamination (see below).

Throughout the work we use the  magnitude relations:

\begin{equation}
\label{eqn1}
M_{B}=M{g}+0.216+0.327\cdot(g-r)
\end{equation}

\begin{equation}
\label{eqn2}
M_{V}=M{g}-0.011-0.587\cdot(g-r)
\end{equation}

\noindent derived from  \citet{CHO08}, where the absolute magnitudes $M_g$ 
were obtained for each galaxy through the distance moduli given in 
\citet{MEI07} and the $(g-r)$ colours from \citet{CHE10}.

From the whole Virgo ACS galaxy sample (100 galaxies), we remove 
galaxies without distance moduli in \citet{MEI07} as well as particular 
subgroups as the ``blue cloud'' or ``red compact'' dwarfs \citep{SMI13}. We also 
rejected galaxies in frames contaminated by the GCs of massive neighbours and 
those that, after removing field contamination, leads to a null content of 
GCs. This procedure leaves a total sample of 67 galaxies.

An estimate of field contamination was carried out by selecting twelve 
galaxies with less than ten globulars and considering as possible field 
interlopers all those objects with galactocentric distances larger than 30 
arcsecs. This statistic, scaled by field size and number of galaxies composing 
a given fiducial galaxy, was subtracted from the composite GC $(g-z)$ colour 
histograms.

Regarding the limiting magnitude that assures an acceptable degree of 
completeness, we compared the number of GCs bluer and redder than $(g-z)=1.2$, 
in 0.5 mag intervals in the $g$ band. The ratio between these numbers for 
the  whole GC sample in \citet{JOR09}, shows a detectable deficit of blue 
globulars for objects fainter than $g=25$ mag, a brightness that we adopt as 
the limit of our statistic. Under the assumption that GCs have fully Gaussian 
integrated luminosity functions, with a mean value $g\approx23.85$, that limit 
leaves out some 15  percent of the total cluster population. 

Finally, we are left with a total of 5470 GCs in 67 galaxies that define 
nine fiducial galaxies. The first fiducials, due to their brightness and 
well populated GCS, contain 5, 3 and 5 galaxies, respectively. In particular, 
the two first composite galaxies appear above the magnitude gap at 
$M_V\approx -20.5$ detectable in the colour magnitude diagram (see, 
for example, \citealt{SMI13}).

In what follows, the interstellar reddening corrections were derived by 
adopting $E(g-z)=$2.15~$E(B-V)$ and the colour excesses listed by 
\citet{FER06} which, in turn, are based on the \citet{SCH98} maps.

\begin{figure}
\includegraphics[width=\hsize]{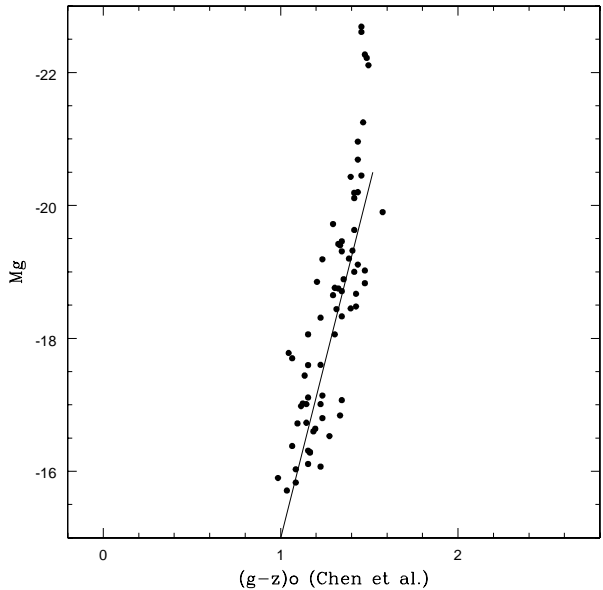}
\caption{Absolute magnitude $M_g$ versus $(g-z)_0$ colour for the selected 
Virgo ACS sample including 67 galaxies. The straight line has the slope of the 
relation found by \citet{SMI13} (see text).}
\label{Mg_gz1}
\end{figure}

The absolute magnitude $M_g$ versus intrinsic colour $(g-z)_0$ relation for the 
galaxy sample, corresponding to a mean colour excess $E(g-z)=0.065$ and mean 
interstellar extinction $A_g=0.050$, is shown in Figure\,\ref{Mg_gz1}, where 
the straight line has the same slope of the fit presented by \citet{SMI13} in 
their study of the colour-magnitude relation of the ACS Virgo cluster galaxies. 
The galaxy members (identified by their VCC numbers) that define a given 
fiducial galaxy, as well as their composite absolute  $M_g$ magnitudes,  
$(g-z)$ colours and average effective radii (computed from the data given by 
table 2 in \citealt{CHE10}) are presented in Table\,\ref{Table_1}.


\begin{table}
\caption{Parameters of the fiducial galaxies}
\setlength\tabcolsep{1.50mm} 
\begin{tabular}{@{}ccccccc@{}}                                 
\hline
 Fiducial & $\rm N_{gal}$  &  $M_g$  & $(g-z)_0$ & $(g-z)_{mod}$(*) &  $M_V$  &  $\rm r_{eff}$ (kpc)\\
\hline
    1   &   5  &  -22.40 & 1.48  &   1.45    &    -22.86  & 13.1\\
    2   &   3  &  -20.99 & 1.46  &   1.41    &    -21.44  &  4.7\\
    3   &   5  &  -20.28 & 1.45  &   1.39    &    -20.73  &  2.5\\
    4   &   9  &  -19.51 & 1.41  &   1.39    &    -19.95  &  1.4\\
    5   &  10  &  -18.92 & 1.37  &   1.37    &    -19.35  &  1.2\\
    6   &  10  &  -18.35 & 1.31  &   1.29    &    -18.77  &  1.7\\
    7   &  10  &  -17.30 & 1.16  &   1.22    &    -17.68  &  1.7\\
    8   &   9  &  -16.70 & 1.19  &   1.14    &    -17.09  &  0.9\\
    9   &   8  &  -16.10 & 1.13  &   1.12    &    -16.47  &  1.1\\
\hline
\label{Table_1}
\end{tabular}

 Galaxies in each of the nine fiducial galaxies identified by their
 VCC number. 
 1) 1226, 881, 763, 1316, 1978;
 2) 1632, 1903, 1231;
 3) 1154, 2092, 759, 1062, 1030;
 4) 1692, 1938, 1664, 944, 1720, 654, 1279, 2000, 778;
 5) 1883, 1242, 355, 784, 1619, 1250, 369, 1146, 1303, 828;
 6) 1630, 698, 1537, 1913, 1321, 1283, 1475, 1178, 1261;
 7) 9, 1087, 1422, 437, 1861, 140, 1910, 856, 543, 1355;
 8) 2019, 1431, 1528, 1833, 200, 1440, 751, 1545, 1049;
 9) 1075, 1828, 1407, 1512, 1185, 2050, 1826, 230;

* $(g-z)$ colours in the scale of \citet{CHE10}; $(g-z)$ model colours 
+0.13 mag.

\end{table}

\begin{figure}
\includegraphics[width=\hsize]{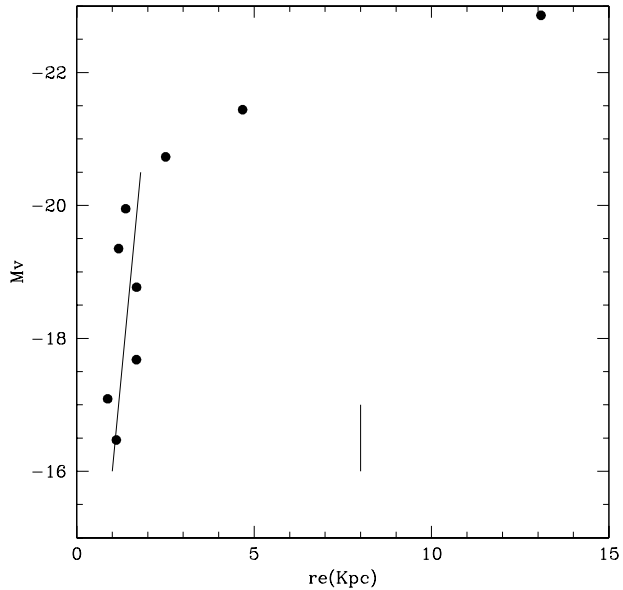}
\caption{Absolute magnitude $M_V$ versus effective radii for nine fiducial 
galaxies resulting from composing the Virgo ACS galaxies shown in 
Figure\,\ref{Mg_gz1}. The short vertical line indicates a radius of 
8 kpc, 100 arcsecs at the distance of the Virgo cluster (adopting a mean 
distance moduli $(m-M)_0=31.1$).}
\label{Mv_rekpc}
\end{figure}

In the case of the brightest galaxies, the areal coverage of the Virgo ACS is 
not large enough for a complete analysis of their GCS. This is illustrated in 
Figure \,\ref{Mv_rekpc} where the short vertical line indicates the size 
(in kpc) corresponding to 100 arcsecs at the mean  distance of the Virgo 
cluster (see also, \citealp{SMI13}).

In order to estimate the total number of GCs in each fiducial galaxy we used 
the data given in HHA13 (their table 4).  First, we grouped the galaxies 
defining each fiducial and then made a regression of the number of GCs in the 
ACS galaxy sample versus those coming from that paper. This analysis indicates 
that significant corrections to the number of GCs are only required for the 
three brightest fiducial galaxies.

A further step for these galaxies is a tentative estimate of the distribution 
of their GCs in terms of the blue and red sub-populations, a 
matter complicated by the different spatial scale lenghts of these populations. 
That is, red GCs are usually more concentrated towards the galaxy centres 
than their blue counterparts, leading to a change of the GCs number ratio 
along the galactocentric radius. 

In these cases, and for illustrative purposes, we take the giant ellipticals 
NGC\,1399 and NGC\,4486, for which wide field studies are available (see, for 
example, FFG07). In the first galaxy the GC sub-population splits by 
$\approx50$ percent, while for NGC\,4486 the blue and red GC 
populations follow a proportion of 70 percent of blue and 30 percent of 
red GCs. As a compromise, we adopt indicative values of 65 and 35 
percent, respectively.

The corrections inferred for the blue and red populations of the 
three brightest fiducials galaxies are given in Table\,\ref{Table_2}. We warn 
the reader about the uncertainties involved (and also recall the inherent 
comments in HHA13). The last two columns in this table give the mean number of 
blue and red GCs after correcting for areal completeness.
  

%
\section{Globular cluster colour distributions}
\label{Color}

The basic ideas for modelling the GC colour histograms were previously 
explained in FFG07, FVF09 and FVF12, and assume the existence of two distinct 
GC sub-populations: the so called blue and red GCs. Note that this is a matter 
of working nomenclature as the red GCs are in fact redder (but not much in 
some of the low mass galaxies) than those that define the omnipresent blue GC 
sub-population. 

As explained in those papers, we use a Monte Carlo generator to create seed 
GCs with a distribution in chemical composition $Z$  controlled by a scale 
length $Z_s$ (measured in $Z_\odot$ units in what follows and listed in 
Table\,\ref{Table_2}), and defined between a lowest value $Z_i$  and an upper 
limit $Z_{sup}$. For both sub-populations we adopt $Z_i=0.014~Z_{\odot}$ (the 
lowest value in the colour-metallicity calibration), while for $Z_{sup}$ we 
adopt 1.0 and 2.3 $Z_\odot$ for the blue and red GCs, respectively. The 
lowest chemical abundance corresponds to $\rm[Fe/H]\approx-2.2$ adopting an 
$\alpha$ ratio in the range of 0.3 to 0.4.
  
The simplest function (i.e. involving a minimum number of parameters) 
that provides a good fit to the colour histograms of a GC population with 
$N_0$ members is:

\begin{equation}
\label{eqn3}
(dN/dZ)= N_0~ exp[-(Z-Z_i)/Z_s],
\end{equation}

\noindent and with our adopted colour-abundance relation, requires 
two metallicity distribution functions (MDFs) identified as blue and red GCs.  
It is worth mentioning that two MDFs for halo stars and GCs, were also adopted 
by \citet{VAN04} and \citet{HAR07}, respectively.

In our approximation, both GC sub-populations, as mentioned before, exhibit 
some degree of overlapping in metallicity (depending on the $Z_s$ parameters). 
This has similarities with the situation reported by \citet{LEA13} for Milky
Way clusters. These authors find a ``bifurcated'' age metallicity relation, 
i.e., the presence of two sequences sharing the same age spread 
($\approx 2~Gy$), shifted in mean metallicity, and exhibiting a range where 
both (halo and disc) GC populations share the same metallicity. As discussed 
in $F13$, such an age spread would be undetectable by photometric means.  

The GC colour histogram fits were performed in a similar way for the blue and 
red clusters except that, for the first sub-population, we added an empirical 
value $\Delta_{Z}=0.65(g-22.35)$ to their $Z_{sb}$ in an attempt to mimic the 
mass-metallicity relation observed for these clusters (see, for example, 
\citealt{HAR09}, \citealt{MIE06}, \citealt{MIE10}). This parameter is set to 
zero for GCs fainter than $g=22.35$.

The logarithm of a given $Z$ value was in turn transformed to integrated 
$(g-z)'$ colour using the colour-metallicity calibration by \citet{USH12}. 
This calibration was linked to other $griz'$ colours as presented in $F13$. 
According to this last work, the GCs $(g-z)'$ colours are connected to 
$(g-z)_{ACS}$ as given in \citet{JOR09} by:

\begin{equation}
\label{eqn4}
(g-z)_{ACS}=(g-z)'+0.07
\end{equation}

\noindent In this modelling we also generate apparent $g$ magnitudes adopting 
fully Gaussian integrated luminosity functions and the parameters (mean 
absolute magnitudes and GCs luminosity function dispersions) given by 
\citet{VILL10}. These magnitudes, in turn, were used for modelling the $(g-z)$ 
colour errors, as given in table 4 of \citet{JOR09}, that were added to the 
model colours.  

The parameters that deliver the best fits to the (composite) observed 
$(g-z)_0$ histograms are listed in Table\,\ref{Table_2}. The number of GCs 
and the $Z_s$ scales of each sub-population, were iterated aiming at 
minimizing the (inverse) quality fit indicator $\chi ^{2}$ defined by 
\citet{COT98}. Typical errors, estimated as in $F13$ are $\pm 0.01$ in 
$Z_{sb}$, $\pm 0.05$ in $Z_{sr}$ and $\pm$ 0.1 in the ratio of blue to red 
GCs.

We note that the less massive fiducial (number 9), still has a detectable 
number of ``red'' GCs that we identify as a bulge-like component. For illustrative purposes, we also give the best fit 
corresponding to a single GC population of blue GCs. Both data points will be 
shown linked by a line in a number of diagrams that follow.
  
\begin{figure*}
\includegraphics[width=\hsize]{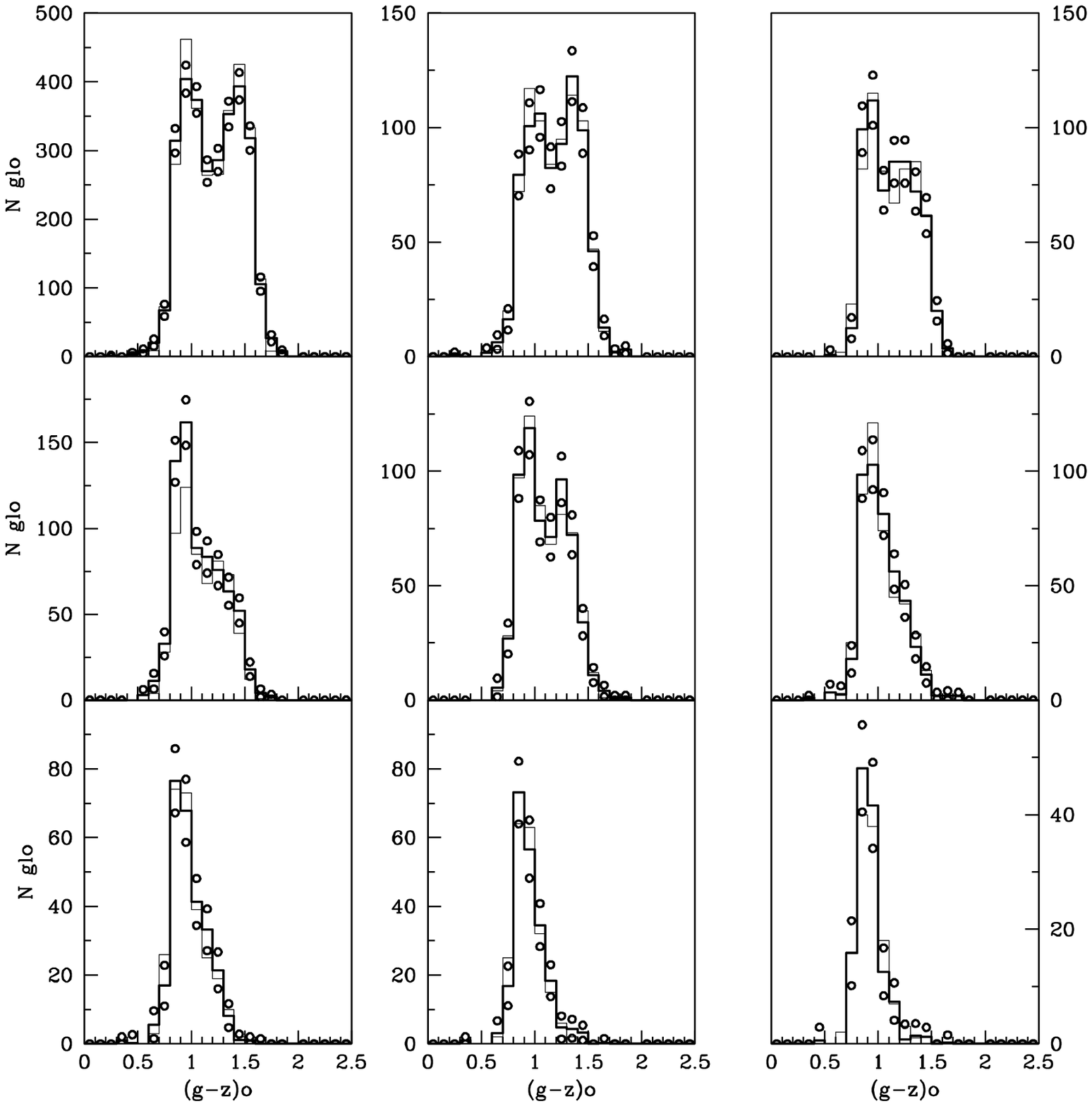}
\caption{Observed and model $(g-z)$ colour histograms for globular clusters in 
the nine composite galaxies (black and grey lines, respectively). The open 
dots represent the formal counting uncertainty in each (0.1 mag) colour bin. 
The absolute $g$ brightness of each fiducial galaxy decreases to the right 
and down.}
\label{histogzall}
\end{figure*}

The GC colour histograms are depicted in Fig.\,\ref{histogzall}, where the 
panels are presented from left to right and down (in order of decreasing 
galaxy brightness) and show a feature already noticed by, \citet{PEN06},
i.e., that bi-modality becomes less evident with decreasing galaxy brightness 
and disappears for the faintest galaxies.  While the blue GC component is 
present in all the galaxies, the red GCs increase their relative importance in 
number, as well as their chemical scale length, as the galaxy mass increases. 
In what follows, we identify the blue GCs with the diffuse low-metallicity 
stellar population, the ``halo'', and the red GCs with the ``bulge-like'' 
stellar population.

\begin{table*}
\centering
\begin{minipage}{150mm}
\caption{Fit parameters for the GC colour histograms}
\setlength\tabcolsep{1.50mm} 
\begin{tabular}{@{}cccccccccc@{}}
\hline
Fiducial &  $N_b$ &   $Z_{sb}$  &   $N_r$  &  $Z_{sr}$  & {$\Delta log(N_b)$} & {$\Delta log(N_r)$} & {$\chi ^2$} & $log\langle N_b \rangle_c$ & $log\langle N_r \rangle_c$\\
\hline
    1   &  1100 &   0.05  &  1852 &  1.10 &   1.17   &   0.54    &   0.98 & 3.515 & 3.304\\
    2   &   260 &   0.05  &   511 &  0.70 &   0.76   &   0.37    &   0.62 & 2.530 & 2.486\\
    3   &   240 &   0.035 &   383 &  0.55 &   0.24   &   0.10    &   1.26 & 1.919 & 1.982\\
    4   &   400 &   0.030 &   373 &  0.55 &    $-$   &   $-$     &   1.01 & 1.643 & 1.613\\
    5   &   260 &   0.030 &   353 &  0.45 &    $-$   &   $-$     &   0.44 & 1.415 & 1.544\\
    6   &   270 &   0.035 &   175 &  0.35 &    $-$   &   $-$     &   1.09 & 1.431 & 1.230\\
    7   &   180 &   0.020 &    92 &  0.25 &    $-$   &   $-$     &   0.48 & 1.255 & 0.954\\
    8   &   160 &   0.020 &    50 &  0.15 &    $-$   &   $-$     &   0.63 & 1.255 & 0.778\\
    9   &   100 &   0.020 &    24 &  0.15 &    $-$   &   $-$     &   0.59 & 1.096 & 0.477\\
    9b  &   124 &   0.020 &     0 &  $-$  &    $-$   &   $-$     &   1.00 & 1.193 &  $-$ \\
\hline
\label{Table_2}
\end{tabular}

$\langle N_b \rangle_c$: mean corrected number of blue GCs.\\
$\langle N_r\rangle_c$:  mean corrected number of red GCs.\\
9b: best fit solution assuming a single blue GC population.
\end{minipage}
\end{table*}


\begin{figure}
\includegraphics[width=\hsize]{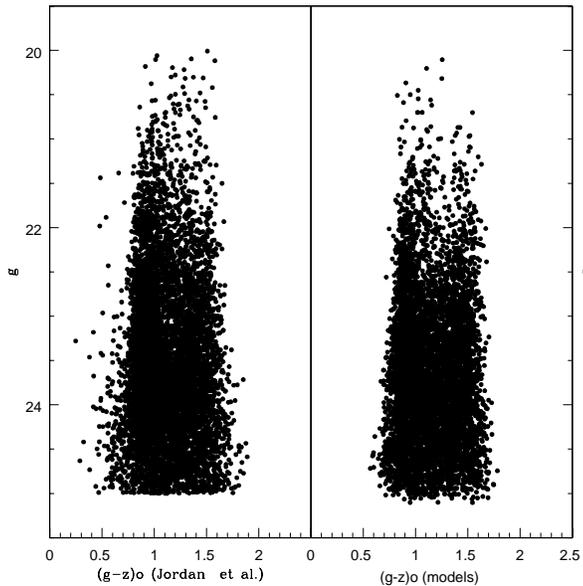}
\caption{{\it Left panel:} Composite colour-magnitude diagram for 5467 
globular clusters associated with the 67 galaxies discussed in the text, 
adopting a limiting magnitude $g=25.0$ mag. {\it Right panel:} Composite 
diagram showing the model globular clusters that deliver the best fits to the 
$(g-z)_0$ colour histograms of globular clusters in the fiducial galaxies. 
The origin of the differences seen in both panels are commented in the text.}
\label{Virgo}
\end{figure}

The composite observed GC colour-magnitude diagram, and the modelled ones, for 
all the fiducial galaxies are shown in Fig.\,\ref{Virgo}. A comparison shows 
an acceptable similarity although some differences exist. On one side, the 
observed sample exhibits a number of objects bluer than $(g-z)_0=0.70$, 
possibly a combination of field interlopers and larger photometric errors. The 
upper bright region of the observed red GCs domain is clearly more populated 
than the model. In this case, we speculate that these objects might be 
GC-ultra compact dwarf transition objects (see, for example, \citealt{FAI11}).


\section{The GC-field stars connection}
\label{FIELD}

In previous works, we attempted an empirical connection between the number of 
GCs and the mass of field stars: 

\begin{equation}
\label{eqn5}
d(N_{gc})/dM_{*}= \Gamma ~ exp[-(\delta~[Z/H])]
\end{equation}

\noindent where both $\Gamma$ and $\delta$ are constant parameters. This 
relation assumes that the number of GCs per unit mass of the associated 
diffuse population increases with decreasing metallicity, and it is consistent 
with the results presented by \citet{HAR02}. These authors found that 
the GCs specific frequency $S_n$ in NGC\,5128 is higher at lower metallicities.
 
The basic idea behind Equation\,\ref{eqn5} is that each GC is formed within a 
much more massive  stellar ``unit''. Each unit shares the colour of the GC 
and its $(M/L)$  ratio is dependent on the chemical abundance $[Z/H]$. In 
particular, the  parameter $\delta$ controls the mass of field stars per GC at 
a given $[Z/H]$ and, after integrating over the whole metallicity range, 
determines the overall $(M/L)$ ratio and integrated colours of a given galaxy. 
In turn, a proper value of the $\Gamma$ parameter leads to its integrated 
absolute magnitude ($M_g$, in our case).
This relation has been successful in providing a good approximation to the 
shape of the brightness profile and colour gradient of the giant elliptical 
NGC\,4486 over a galactocentric radius of 1000 arcsecs (FVF12). 

For the diffuse stellar units within a galaxy, and as in previous 
works,  we adopted:

\begin{equation}
\label{eqn6}
(M/L)_B= 3.71 + ([Z/H] + 2)^{2.5}
\end{equation}

\noindent that gives a good representation for mass to luminosity ratios in the
blue band, between $[Z/H]=$ -2 and 0.5 as given by \citet{WOR94}. For lower 
and higher metallicities we assume $(M/L)_{B}=$ 3.70 and 13.6, respectively.

In this approach, the integrated colours of the ETGs will be determined by the 
luminosity weighted mass-metallicity spectrum of the stellar units that 
represent field stars. Without obvious star forming processes, these colours 
should be confined to the GCs colour domain (see below).

\section{Photometric scales for globular clusters and galaxies}
\label{ZEROPOINTS}

As previously shown in FVF09, a comparison between the GCs $(g-z)_{ACS}$ 
colours presented in \citet{PEN06} (in the form of GC colour histograms) and 
those of the galaxies, from \citet{FER06}, suggests a zero point difference 
between both data sets that is independent of magnitude. The $(g-z)$ galaxy 
colours in this last work, in turn, are very similar to the colours given in 
the multicolour photometry presented by \citet{CHE10}, that we adopted in this 
paper.

In this section we revise the connection between the photometric scales 
through two different pathways:

\begin{description}
\item[{\it a)}] By analysing the galaxy-GCs colour difference in two galaxies 
for which we have independent photometric data sets: NGC\,4649 and NGC\,4486.
\item[{\it b)}] Using the results of our galaxy colour modelling for ten of 
the brightest galaxies in the sample.
\end{description}

\citet{GOU13} noted that there is a significant colour difference between the 
red GCs (defined as having $(g-z)_{ACS}>1.2$) and the galaxy colours, six of 
them included in our analysis. These authors suggest that the difference 
arises as a consequence of a ``bottom heavy'' luminosity function for the field 
stars in these massive galaxies that would be different from that prevailing 
in GCs. We return to this subject later in this section.

\subsection{The galaxy-GCs colours in NGC\,4486 and NGC\,4649} 

Aiming at decreasing the effects of incompleteness in the GC samples, due to 
the bright inner regions of the galaxies, we restrict our analysis to the 
galactocentric range defined between 50 and 100 arcsecs, and to clusters 
brighter than the turn-overs of the integrated luminosity function at 
$g\approx23.85$ (\citealt{VILL10}).

In the case of NGC\,4486 we use the $(C-T_1)$ GCs photometry presented in F13. 
There is no $(C-T_1)$ colour profile available for this galaxy and, then, we 
adopted the $(B-V)$ surface photometry given by \citet{ZEI93}. This 
photometry, in very good agreement with the $(B-V)$ colours obtained by 
extrapolating (inwards) the colour profile presented by \citet{RUD10}, gives 
$(B-V)_0=0.93$.

In order to link the $(C-T_1)$ and $(B-V)$ colours, we use an empirical 
relation derived from Milky Way GCs, adopting the data given by 
\citet{HARC77} and \citet{REE88}, and the interstellar colour excesses from
\citet{REC05} when available, or from Reed et al. otherwise. For 47 objects 
in common in those works, we performed a bisector least-square fit that leads 
to:

\begin{equation}
\label{eqn7}
(C-T_1)_0= 2.12~(\pm0.08)~(B-V)_0 - 0.21~(\pm0.01)
\end{equation}

\begin{equation}
\label{eqn8}
(B-V)_0 = 0.47~(\pm0.02)~(C-T_1)_0 + 0.10~(\pm0.01)
\end{equation}

For 458 GCs in NGC\,4486, we obtained a mean colour $(C-T_1)_0=1.51$. 
From these last relations, $(B-V)_0=0.81$ and then galaxy-GC colour differences $\Delta(B-V)=0.13$
 and $\Delta(C-T_1)=0.28$. In turn, from the colour-colour relations in F13, we infer $\Delta(g-z)=0.24$.

Alternatively, the GCs photometry from \citet{JOR09} yields a mean 
$(g-z)_0=1.16$, for 435 clusters brighter than the turn-over at $g=23.85$, 
while the galaxy photometry by \citet{CHE10}, gives $(g-z)_0=1.54$, and then, 
$\Delta(g-z)=0.39$. This last colour difference results 0.15 mag larger than 
our previous estimate.\\ 
  
\citet{FAI11} presented $gri'$ Gemini photometry for GCs in NGC\,4649, and for 
the inner region of the galaxy. The mean $(g-i)'_0$ colour of the GCs results 
$0.94\pm 0.03$, while the galaxy halo colour is $(g-i)'_0=1.08$ (see their 
figure\,10). These values lead to a colour difference between the galaxy and
its GCs of $\Delta(g-i)'_0= 0.14$ mag, and to $\Delta(g-z)= 0.20$ mag from 
the relations in \citealt{FOR13}. 

In turn, the $(g-z)_{ACS}$ photometry by \citet{JOR09} for 207 GCs, yields 
a mean colour $(g-z)_0=1.20$, while the photometry given by \citet{CHE10} 
indicates $(g-z)_C=1.52$, then leading to $\Delta(g-z)=0.32$ mag. That is, 
0.12 mag larger than our estimate.

In what follows, we adopt a colour zero point difference between the (galaxy) 
photometry of \citet{CHE10}/\citet{FER06} and the (GCs) photometry
by \citet{JOR09}, of $+0.13$ mag.

We cannot asses the origin of such a difference but note that \citet{JAN09} 
have also pointed out that the $(g-z)$ galaxy colours by \citet{FER06} are 
sistematically redder by about 0.1 mag compared with their own photometry.
     
\subsection{Colour modelling for ten bright galaxies}
    
We selected the ten brightest galaxies in the Virgo ACS sample after excluding 
VCC\,798, which presents a blue central region indicating recent star 
formation, and VCC\,1535 and VCC\,2095 which are disky edge on galaxies. 
For these ten galaxies we attempted our colour modelling, first by fitting the 
GC $(g-z)_{ACS}$ colour histograms (based on the \citealt{JOR09} photometry) 
and then, looking for a  single $\delta$ parameter that leads to the 
integrated galaxy colours. For this approach, and given the large effective 
radii of these galaxies (comparable or larger than the ACS field, see 
\citealt{SMI13}) we adopted the colours within 1 $\rm r_{eff}$ as given by 
\citet{CHE10}.

The results corresponding to $\delta=1.9$, and adding $0.13$ mag to our model 
colours, are listed in Table\,\ref{Table_3}. The mean residual in the $(g-z)$ 
colours is $0.015 \pm 0.038$ mag, i.e, the models match the observed colours 
with a dispersion comparable to the photometric errors for the galaxies.

As a consequence of this last analysis, we assume the $(g-z)'-(g-z)_{ACS}$ 
relation given in F13 and add 0.13 mag to the model colours in order to keep 
the galaxy-photometric scales of \citet{CHE10} and \citet{FER06}. 

\begin{table*}
\centering
\caption{Model parameters for ten bright galaxies of the ACS Virgo Cluster 
Survey.}
\begin{tabular}{@{}ccccccccc@{}}                                 
\hline
       NGC &  VCC &  $N_b$ &  $Z_{sb}$ &  $N_r$ &  $Z_{sr}$ & $\chi ^{2}$  &  $(g-z)_0$ &  $(O-C)_{(g-z)}$ \\
\hline 
      4472 & 1226 & 170 & 0.05 & 330 & 1.20 & 0.83  &  1.51  & $-0.02$\\ 
      4486 & 1316 & 400 & 0.05 & 838 & 1.15 & 1.19  &  1.55  & $+0.01$\\   
      4649 & 1978 & 200 & 0.06 & 385 & 1.30 & 0.69  &  1.52  & $-0.02$\\   
      4406 & 881  & 115 & 0.03 & 144 & 0.40 & 0.88  &  1.42  & $+0.07$\\
      4374 &  763 & 174 & 0.03 & 189 & 0.60 & 0.70  &  1.41  & $+0.00$\\
      4365 &  731 & 170 & 0.03 & 414 & 0.55 & 0.98  &  1.50  & $+0.07$\\    
      4552 & 1632 & 100 & 0.05 & 229 & 0.80 & 0.23  &  1.50  & $+0.00$\\    
      4621 & 1903 &  60 & 0.03 & 186 & 0.60 & 0.42  &  1.45  & $-0.01$\\  
      4473 & 1231 &  87 & 0.04 & 110 & 0.60 & 0.43  &  1.42  & $-0.01$\\
      4459 & 1154 &  60 & 0.03 &  90 & 0.50 & 1.08  &  1.46  & $+0.07$\\
\hline 
\label{Table_3}
\end{tabular}

$(g-z)_0$: Redenning corrected galaxy colours from \citet{CHE10}, within 
1 $\rm r_{eff}$.
\end{table*}

The adoption of a colour difference $(g-z)-(g-z)_{ACS}=0.13$ mag, constant 
with galaxy magnitude, and a mean colour excess $E(g-z)=0.065$, leads to the 
following connection between the $(i-z)'$, $(g-i)'$ and $(g-z)'$ intrinsic 
GC colours from F13, and the observed galaxy colours given in \citet{CHE10}:

\begin{equation}
\label{eqn9}
(i-z)' = (i-z)-0.12
\end{equation}

\begin{equation}
\label{eqn10}
(g-i)' = (g-i)-0.15
\end{equation}

\begin{equation}
\label{eqn 11}
(g-z)' = (g-z)-0.27
\end{equation}

\begin{figure}
\includegraphics[width=\hsize]{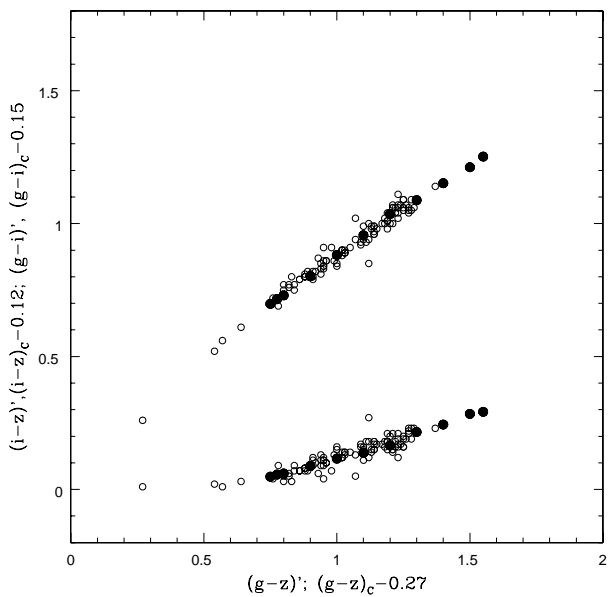}
\caption{Colour-colour diagrams for CGs (large filled dots) from \citet{FOR13} 
and galaxies (small open dots) from \citet{CHE10} (shifted as 
indicated in the text). {\it Upper relation:} $(g-i)$ versus $(g-z)$. 
{\it Lower relation:} $(i-z)$ versus $(g-z)$.}
\label{iz_gi_gz}
\end{figure}

The GCs and galaxy colours, adopting these last relations, are displayed in 
Fig.\,\ref{iz_gi_gz}. We emphasize two aspects connected with this diagram. 
First, that the $(g-i)'$ and $(g-z)'$ GC colours (from a field in NGC\,4486) 
are in excellent agreement with the photometry given by \citet{SIN10} for GCs 
in NGC\,5128. Second, that the galaxies follow colour-colour relations with 
slopes similar to those of the GCs. This result is expected if, in fact, the 
dominant stellar populations in ETGs are traced by the clusters. 

The red GCs in the inner regions of NGC\,4486 show a peak at $(g-z)_{ACS}=1.40$
(or $(g-z)'=1.33$; see FVF12) indicating that these clusters are as red as the 
reddest galaxies in Fig.\,\ref{iz_gi_gz}. That is, we do not find a 
significant offset between the colours of the red GCs and the colours of the 
galaxies as found by \citet{GOU13}. In addition, the bluest galaxies in this 
figure are bluer than the GCs with the lowest metal abundance. All of them 
show blue colour gradients towards their centres indicating recent star 
forming processes.


\section{Parameters inferred from the galaxy colour and brightness fits}
\label{COLOURS}
     
We start by iterating the $\delta$ parameter until the model matches the  
composite integrated colours of the fiducial galaxies within one effective 
radius. A comparison of these (reddenning corrected) colours with those 
corresponding to the whole galaxy does not show significant differences.  

Then, $\Gamma$ was changed until the absolute $M_g$ magnitude of the models 
coincide with the observed ones. Remarkably, the  mean $\delta$ for all 
galaxies  exhibits a relatively low dispersion around the mean values 
($\delta= 1.92\pm0.13$), very similar to that derived by FVF12 in their 
analysis of the galactocentric colour gradient observed in NGC\,4486
($\delta=1.8$). In turn, $\Gamma$ is also almost constant for galaxies fainter 
than the ``gap'' in the colour magnitude diagram, yielding a mean value 
$\Gamma= 0.72\pm0.16$; in $10 ^{-9}$ units). Instead, the two brightest fiducials
 show an increase of this parameter with stellar mass as shown in Table\,\ref{Table_4}.

For the three brightest fiducials, we explored different models by changing 
both the $Z_{sb}$ and $Z_{sr}$ parameters, and the ratio of blue to red GCs. 
The most significant change is that, adopting an equal number of blue and red 
GCs outside the areas covered by the ACS field, the mass of the low metallicity 
halos decreases between 30 percent and 15 percent.

\begin{figure}
\includegraphics[width=\hsize]{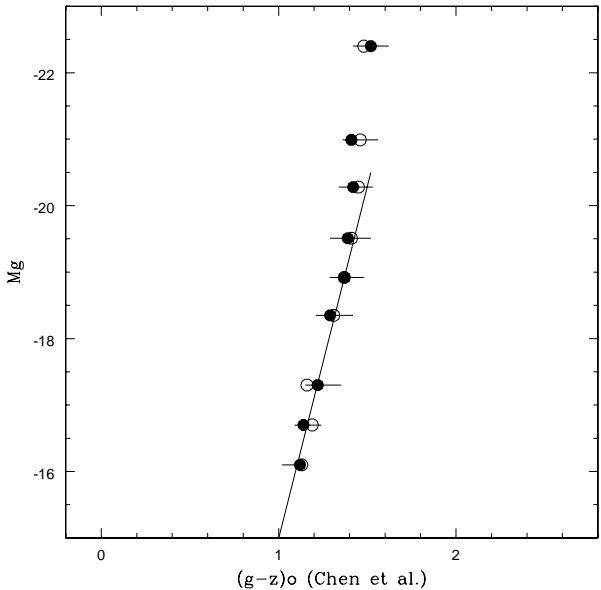}
\caption{Colour-magnitude diagram for the composite fiducial galaxies (open 
dots) compared with the colour modelling explained in the text (filled dots), 
corresponding to a single parameter $\delta=1.9$. The solid line has the same 
meaning as in Figure\,\ref{Mg_gz1}. The overall colour rms resulting from the
observed versus modelled colours is $\pm 0.04$ mag. The horizontal lines show the 
effect of changing the $\delta$ parameter from 0.95 to 3.80.}
\label{Mg_gz2}
\end{figure}

The model colours of the fiducial galaxies corresponding to a single mean parameter 
$\delta=1.9$ are shown in Figure\,\ref{Mg_gz2} (where open dots represent the 
fiducial galaxies and filled dots corresponds to the model fits) and are also 
presented in Table\,\ref{Table_1}.
In this figure we add 0.13 mag to our modelled colours in order to keep the 
same colour scale as in Fig.\,\ref{Mg_gz1}. The mean difference between 
observed and modelled colours for all the fiducials is $-0.012\pm0.04$ mag. We 
stress that this rms value is coherent with the overall photometric errors.
This figure also shows the effect of changing the $\delta$ parameter (as 
horizontal bars) from 0.5 to 2 times the adopted value and correspond to 
$d(g-z)/d(log(\delta)\approx0.18$ to 0.10 (for the most and least massive 
galaxies, respectively). This shallow dependence of the model galaxy colours 
with $\delta$ is a consequence of the shapes of the distribution of the 
integrated stellar brightness with chemical abundance (or colour). For the 
range of $Z_s$ parameters that characterize a given GC population, these 
distributions are broad and bell shaped. A variation of $\delta$ (for a given 
$Z_s$ value) changes the skewness of these distribution, but has a relatively 
low impact on the mean colour. This situation also leads to the need to 
define fiducial galaxies since their mean colours are less affected by 
photometric errors, and also decrease the effect of particular star-forming 
events within a given individual galaxy.  
   

The halo, bulge-like and total composite masses obtained through the modelling 
procedure are given in Table\,\ref{Table_4}.


\begin{table*}
\centering
\begin{minipage}{150mm}
\caption{Model results for the fiducial galaxies}
\begin{tabular}{@{}ccccccccccc@{}}                                 
\hline
   Fiducial & {$\Gamma ^{**}$} & $log(M_{halo})$ & $log(M_{bulge})$ & $log(M^{*})$ & $log(t)$ & $log(t_{b~})$ & $log(t_{r~})$ & $log(Sm)$ & $(M/L)B$ & $[Z/H]$\\
\hline
      1    &  2.44 &    10.984     &    11.757    &  11.850  &  0.853  &    1.695  &   0.800   &  $-$0.448  &  7.6 &  $-$0.15\\
      2    &  1.09 &    10.343     &    11.196    &  11.253  &  0.556  &    1.187  &   0.232   &  $-$0.899  &  7.1 &  $-$0.27\\
      3    &  0.59 &     9.981     &    10.903    &  10.952  &  0.300  &    0.938  &   0.079   &  $-$1.198  &  6.8 &  $-$0.32\\
      4    &  0.53 &     9.760     &    10.586    &  10.646  &  0.283  &    0.883  &   0.027   &  $-$1.179  &  6.8 &  $-$0.32\\
      5    &  0.67 &     9.421     &    10.354    &  10.402  &  0.383  &    0.993  &   0.192   &  $-$1.217  &  6.6 &  $-$0.38\\
      6    &  0.61 &     9.502     &     9.997    &  10.118  &  0.525  &    0.929  &   0.233   &  $-$1.102  &  5.8 &  $-$0.57\\
      7    &  0.76 &     9.121     &     9.518    &   9.664  &  0.767  &    1.134  &   0.436   &  $-$0.902  &  5.2 &  $-$0.73\\
      8    &  0.88 &     9.049     &     9.065    &   9.358  &  1.010  &    1.206  &   0.713   &  $-$0.702  &  4.5 &  $-$1.03\\
      9    &  0.98 &     8.853     &     8.765    &   9.113  &  1.077  &    1.244  &   0.712   &  $-$0.605  &  3.9 &  $-$1.06\\
      9b   &  0.75 &     9.057     &     0.000    &   9.057  &  1.130  &    1.130  &    $-$    &  $-$0.605  &  3.9 &  $-$1.46\\
\hline
\label{Table_4}
\end{tabular}
* Masses derived adopting $\delta=1.9$\\
** {$\Gamma$} in $10 ^{-9}$ units.
\end{minipage}
\end{table*}

\begin{figure}
\includegraphics[width=\hsize]{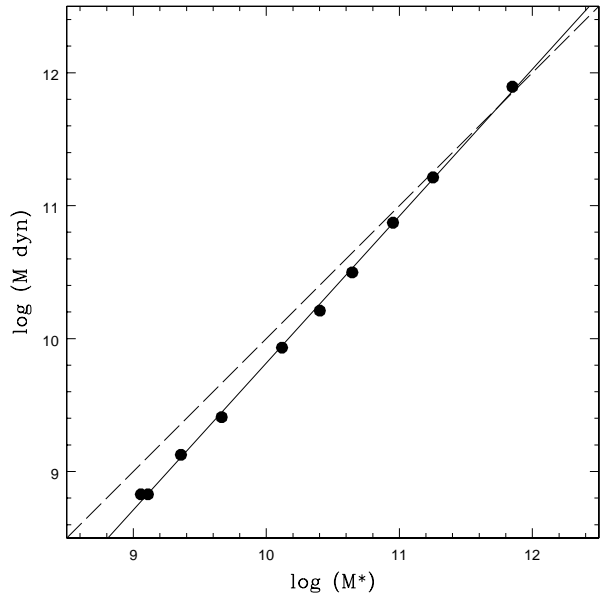}
\caption{Comparison between the total stellar masses derived in this work with 
those coming from the so called {\it dynamical masses} as defined by 
\citet{WOL10} and calibrated in terms of $L_V$ by \citet{HAR13}. The solid 
line corresponds to a least-square fit (see text) and, the dashed one, to a 
1:1 relation.}
\label{logmdy_logmt}
\end{figure}

In Figure\,\ref{logmdy_logmt} we present a comparison of the masses derived in 
this paper and those coming from the stellar mass-$L_V$ calibration presented 
by HHA13 and based on the so called {\it dynamical} masses derived for 
spheroidal systems by \citet{WOL10}. An least-square fit leads to:

\begin{equation}
\label{eqn12}
log(M_{dyn})=1.105~log(M_{*})-1.237
\end{equation}
   
\noindent This equation shows that our model masses become larger than the 
dynamical masses when $M_{*}$ decreases. The largest deviation between both 
calibration amounts to $\Delta log(M_{*})=0.28$ at $log(M_{*})=9.0$.


\begin{figure}
\includegraphics[width=\hsize]{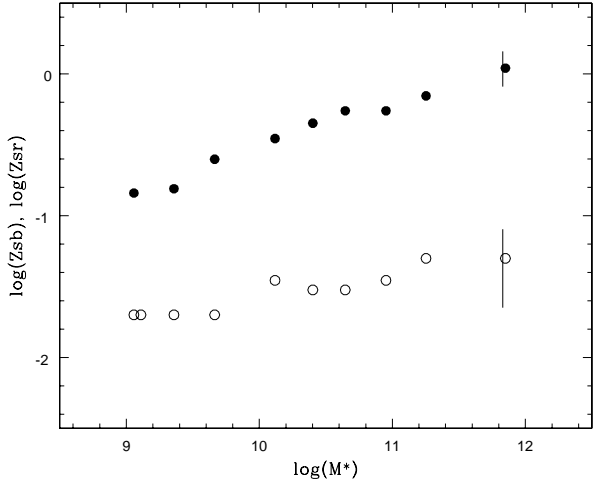}
\caption{Chemical scale lengths corresponding to the red globular clusters 
(filled dots) and blue ones (open dots) as a function of the total stellar 
galaxy mass corresponding to the fiducial galaxies. The scale lengths of the 
three most massive galaxies are upper values derived for the central regions 
of these objects. Note that red clusters appear above $log(M_{*})=9.25$ and 
exhibit a much evident change with stellar mass than their blue counterparts. 
The vertical lines represent the variation of these parameters with 
galactocentric radius in NGC\,4486 (FVF12).}
\label{logZs_logmt}
\end{figure}

The trends of the GC chemical scale lengths $Z_s$ with total stellar mass are 
depicted in Figure\,\ref{logZs_logmt} for each GC sub-population. Red GCs 
exhibit a marked variation in comparison with the more shallow increase 
corresponding to the blue GCs. The $Z_s$ parameters of the three most massive 
fiducials are upper values as they come from the innermost regions of these 
objects. In this figure, and as a reference, we show the range covered by the 
$Z_s$ values of both sub-populations, as a function of galactocentric radii, 
in NGC\,4486 (from FVF12).

The behaviour of the chemical scale length of the blue GCs with galaxy mass 
(or luminosity) is consistent with a fact already pointed out by 
\citet{STRA04} in the sense that these clusters seem ``to know'' about the 
galaxy they are associated with.

\begin{figure}
\includegraphics[width=\hsize]{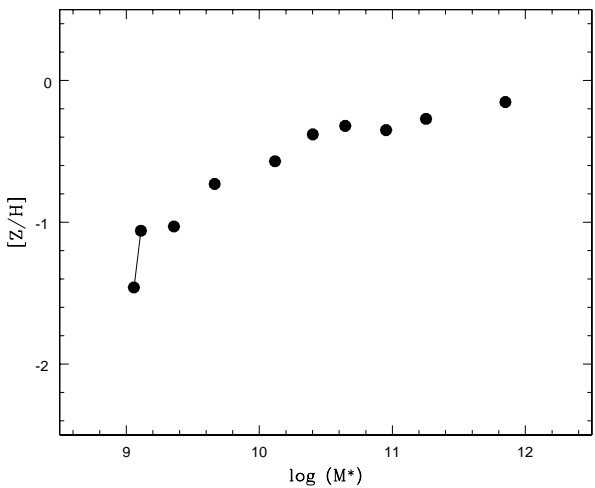}
\caption{Inferred integrated chemical abundance (weighted by B-band luminosity) as a
 function of total stellar mass for each fiducial galaxy. Two possible 
models for the fiducial galaxy number 9 (see text) are shown connected by a 
line.}
\label{ZH_mt}
\end{figure}

The composite (B-band luminosity weighted) chemical composition $[Z/H]$ is 
displayed in Figure\,\ref{ZH_mt}. This diagram shows that a single power law 
does not yield a proper representation over the whole range of stellar mass  
and rather exhibits a smooth roll-over.


\begin{figure}
\includegraphics[width=\hsize]{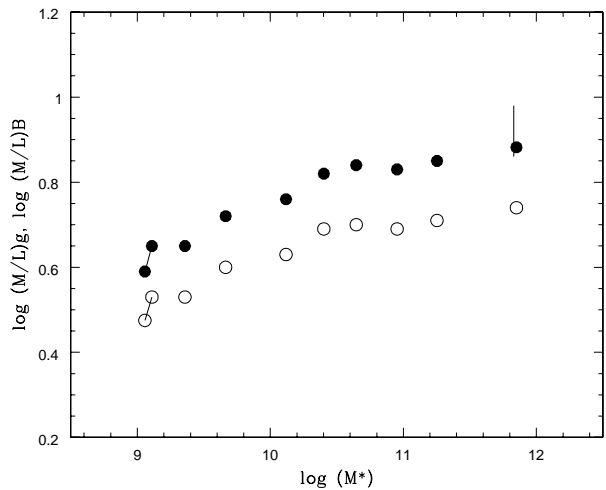}
\caption{Inferred B-band (filled dots) and g-band (open dots) luminosity to 
stellar mass ratios as a function of stellar mass for all the fiducial 
galaxies. The ratios corresponding to the three most massive galaxies represent
upper limits corresponding to the central regions of the galaxies. The 
vertical line at the right shows the galactocentric variation determined by 
\citet{FVF12} in the case of NGC\,4486. }
\label{RMLB}
\end{figure}

A single power law does not seem either an appropriate fit for the variation 
of the  mass to luminosity ratio $(M/L)_{B}$ as seen in Fig.\,\ref{RMLB}. This 
relation is in good agreement with the results presented by \citet{TOR09}, 
through a completely different approach, and also with the mass to 
$g$-luminosity ratios derived by \citet{KAU03} and \citet{BEL03} (see their 
figures 14 and 6, respectively).

\begin{figure}
\includegraphics[width=\hsize]{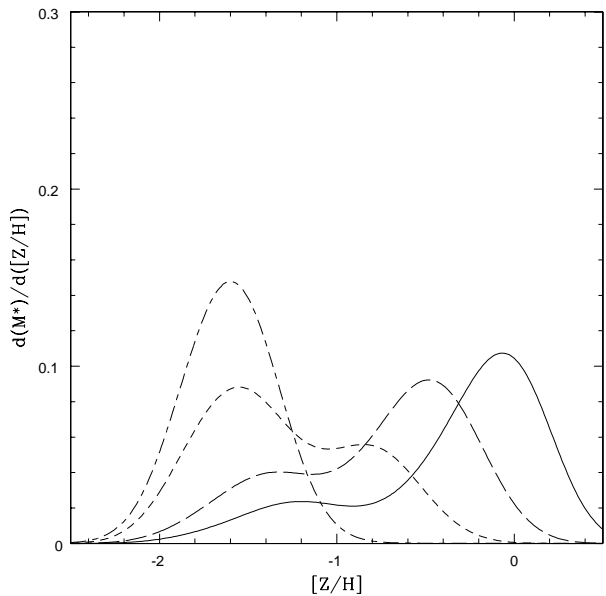}
\caption{Inferred chemical abundance distributions for the stellar 
populations in the fiducial galaxies number 1 (solid line), 6 (long-dashed) 
and 9 (short dashed) normalized by mass and convolved with a Gaussian kernel
($\sigma=$ 0.20 dex in [Z/H]). This figure also includes the 
expected distribution for the case of a bulgeless galaxy comparable to this 
last fiducial (long-short-dashed line).}
\label{ZH_stars}
\end{figure}

As illustrative examples, Fig.\,\ref{ZH_stars} displays the stellar mass 
distribution (normalized by total mass) inferred for the fiducial galaxies 
number 1, 6 and 9 on the basis of the parameters listed in 
Table\,\ref{Table_2}.  Note that, for the most massive object, the stellar 
distribution corresponds to its inner region and resembles the observed one 
for resolved stars in NGC\,5128 \citep{HAR02,REJ11}, i.e., a broad and 
asymmetric high metallicity (bulge-like) component and an extended
low-metallicity tail. In the case of the least massive fiducial galaxy 
(number 9) we show an alternative distribution assuming that this galaxy only 
has blue GCs.

 We use the term ``resembles'' since a more thorough comparison will require 
a detailed analyisis of the metallicity scales (and the different inherent 
errors) adopted in this and other works. In this last figure, the chemical 
abundances of the model have been convolved with a Gaussian  kernel (with 
$\sigma=0.20$ dex) in a preliminary approach to asses the effects of such
errors. For example, these errors smooth-out the eventual presence of 
``valleys'' in the chemical abundance distributions arising as the result of 
combining halo and bulge components.

The integrated colour of the last fiducial (on our photometric scale) is 
$(g-z)_0=0.99$ or $(C-T_1)_0=1.32$ (from the relations given in F13) and, then,
$(B-V)_0=0.72$, which after adding an interstellar colour excess 
$E(B-V)=0.02$, leads to $(B-V)=0.70$, a colour coincident with the photometric 
results given by \citet{MIHHA13} for the outer region of NGC\,4472. This 
coincidence may indicate that galaxies with comparable, or with lower masses, 
can be the principal contributors to the formation of the low-metallicity 
halos. 
     

\section{The number of globular clusters versus galaxy mass relation}
\label{BREAK}
     
\begin{figure}
\includegraphics[width=\hsize]{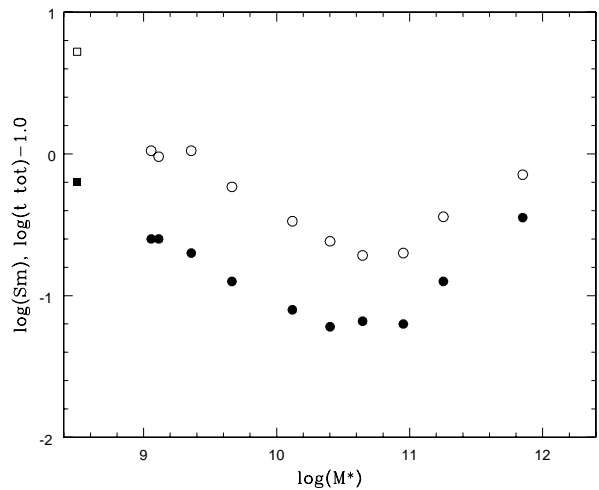}
\caption{Formation efficiency $t_{tot}$ (open dots; shifted by -1) and percent 
mass locked in globular clusters, $S_m$, (as defined in HHA13; filled dots), 
as a function of total stellar mass of each fiducial galaxy. The open and 
filled squares at left come from the HHA13 fit shown in their figure 14.}
\label{Delta_HARRIS}
\end{figure}

The total number of GCs was used to derive the ``GC formation efficiency'' (i.e. 
number of GCs per stellar mass unit), $t$ (in $10^{-9}$ units) and also $S_m$, 
the GC mass fraction content of each galaxy, as defined in HHA13. To derive 
this last parameter we adopted their power law approximation to obtain the 
mean GC mass as a function of the stellar mass of the galaxy. Both $t$ and 
$S_m$ are displayed in Figure\,\ref{Delta_HARRIS}. These parameters exhibit 
trends similar to those obtained by HHA13 (see their figure 14).


\begin{figure}
\includegraphics[width=\hsize]{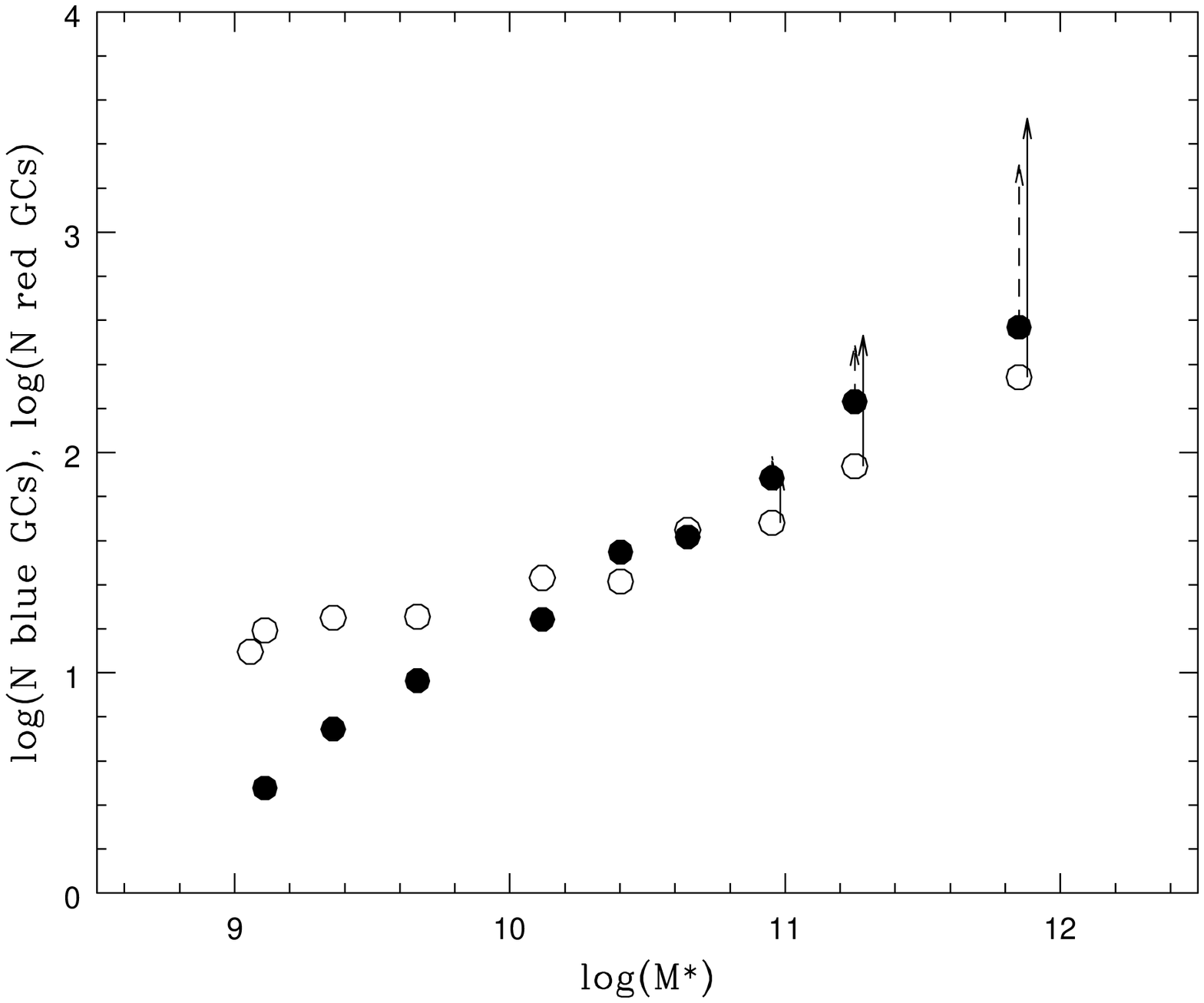}
\caption{Number of blue GCs (open dots; solid lines) and red GCs (filled dots;
 dashed arrows) as a function of total stellar mass of each fiducial galaxy. The
 arrows indicate a tentative correction due to incomplete areal coverage (see text).}
\label{nb_nr_corr_mt}
\end{figure}

The mean number of blue and red GCs for each fiducial galaxy are shown in 
Figure\,\ref{nb_nr_corr_mt}. An interesting feature 
in these diagrams is that below $log(M_{*}) \approx 11.0$,  where completeness 
factors do not play an important role, the slopes that characterize the 
increase of the number of blue and red GCs with total stellar galaxy mass are 
significantly different ($0.30\pm0.03$ and $0.70\pm0.02$, respectively).


\begin{figure}
\includegraphics[width=\hsize]{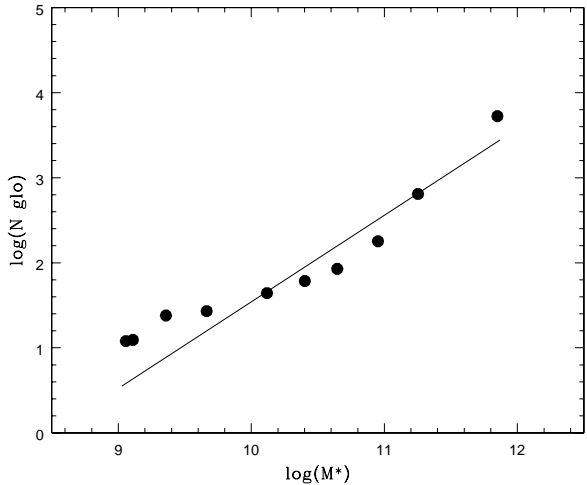}
\caption{Total number of globular clusters (blue and red, added) as a function 
of the fiducial galaxy stellar masses. The straight line has the slope 
determined by \citet{HAR13}. This diagram shows that a linear log-log
approximation, as suggested by these authors, in fact is not a good 
representation of the number of clusters-galaxy stellar mass relation.}
\label{logn_HARRIS}
\end{figure}

Alternatively, combining the number of blue and red GCs leads to the behaviour 
displayed in Figure\,\ref{logn_HARRIS} corresponding to the total GC 
population. This diagram also supports the claim of HHA13 in the sense that
the number of GCs is not a a good indicator of the stellar mass of the galaxy. 
The straight line in this diagram has the same slope of the fit shown by these 
authors in their figure 9. Thus, the behaviour pointed out by HHA13 is 
consistent with the presence of two distinct GC populations.
       

\section{The Sersic index and the projected stellar mass density}
\label{Sersic}

\begin{figure}
\includegraphics[width=\hsize]{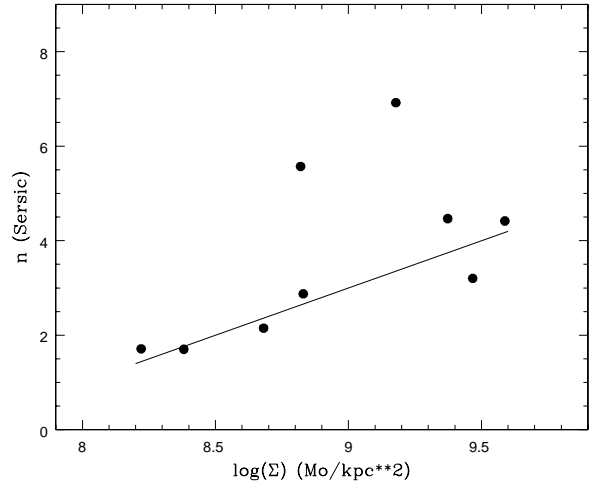}
\caption{S\'ersic $n$ index versus surface stellar density $\Sigma$ (in 
$M_\odot$ per square kpc) obtained as explained in the text. A linear fit 
gives a good approximation for most fiducial galaxies except the two brightest 
ones.}
\label{n_MU2}
\end{figure}

Table\,\ref{Table_5} gives the S\'ersic $n$ index and the surface stellar mass density 
$\Sigma$ (in $M_{\odot}$ per square kpc) for the fiducial galaxies. The first 
parameter is just a mean value for the galaxies contained with each fiducial 
group as given by \citet{CHE10}. In turn, $\Sigma$ was estimated by taking the 
mean surface brightness in the $g$-band within one effective radius (mag per 
square arcsec) (table 2 in \citealt{CHE10}), transforming this magnitude to B 
magnitudes (through the relations given in Section \ref{FIDUCIALS}), and 
averaging all the galaxies within a given fiducial group. Finally, adopting 
the \citet{MEI07} distance moduli, and the mass to B-band luminosity ratios 
given en in Table\,\ref{Table_4}, these surface brightnesses were transformed
to solar masses per square kpc. The relation between $n$ and $log(\Sigma)$ is 
shown in Figure\,\ref{n_MU2}

\begin{table}
\caption{S\'ersic's parameter n and proyected stellar mass density $\Sigma$ (in $M_\odot/Kpc^2$) 
for the fiducial galaxies}
\label{Table_5}
\setlength\tabcolsep{14.00mm} 
\begin{tabular}{@{}ccc@{}}
\hline
           Fiducial  & log(n) &  {$log (\Sigma)$}\\
\hline
             1   &     0.746    &    8.82\\
             2   &     0.840    &    9.18\\
             3   &     0.650    &    9.37\\
             4   &     0.645    &    9.59\\
             5   &     0.505    &    9.47\\
             6   &     0.459    &    8.83\\
             7   &     0.231    &    8.38\\
             8   &     0.332    &    8.68\\
             9   &     0.233    &    8.22\\
\hline
\end{tabular}
\end{table}

Only the two brightest galaxies appear well off, as seen in 
similar diagrams (for example, \citealt{GRA03}). This figure shows an increase 
of $n$ from $\approx1.8$ to $\approx4$, i.e, the fiducial galaxy profiles 
become more de Vaucouleurs-like as the projected stellar mass density 
increases.

\begin{figure}
\includegraphics[width=\hsize]{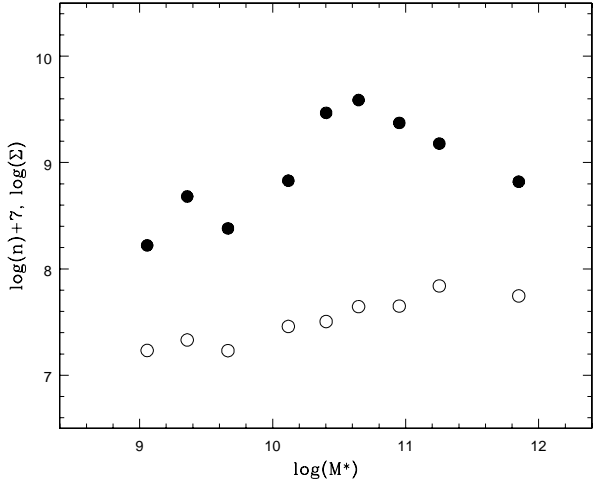}
\caption{Variation of the logarithm of the S\'ersic $n$ index (shifted by 
+7; open dots) and projected stellar mass density with total stellar mass of 
the fiducial galaxies (solid dots). There is a marked peak at 
$log(M_{*})\approx10.5$ and then a decrease with increasing stellar galaxy 
mass.}
\label{n_surf_mt}
\end{figure}

Both $n$ and $\Sigma$, are displayed as a function of total stellar mass (in 
logarithmic format) in Figure\,\ref{n_surf_mt}. This diagram shows the well 
known increase of $n$ with galaxy mass and brightness (see, for example, 
\citealt{DON11}). In turn, the projected surface stellar mass increases until 
reaching a well defined peak at $log(M_{*})\approx 10.5$ and an almost 
symmetric decrease for the brightest fiducials (i.e., for galaxies above the 
gap in the colour-magnitude diagram). This diagram is in fact a transformed 
version of figure\,1 in \citet{KOR09}, and the peak of the $\Sigma$ value
would be the boundary that, according to these authors, indicates a 
``dichotomy'' in the galaxy forming process.


\section{Halo and Bulge Masses and their globular cluster formation 
efficiencies}
\label{HALOBULGE}

\begin{figure}
\includegraphics[width=\hsize]{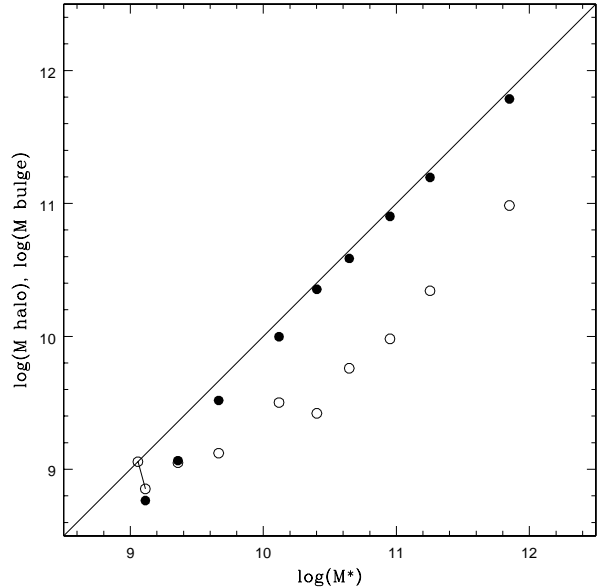}
\caption{Variation of the bulge stellar mass associated with the red globulars 
(filled dots), and that corresponding to the low-metallicity stellar halo 
connected with the blue globulars (open dots). The straight line is the 1:1 
relation.}
\label{mhalo_mbulge_mt}
\end{figure}

\begin{figure}
\includegraphics[width=\hsize]{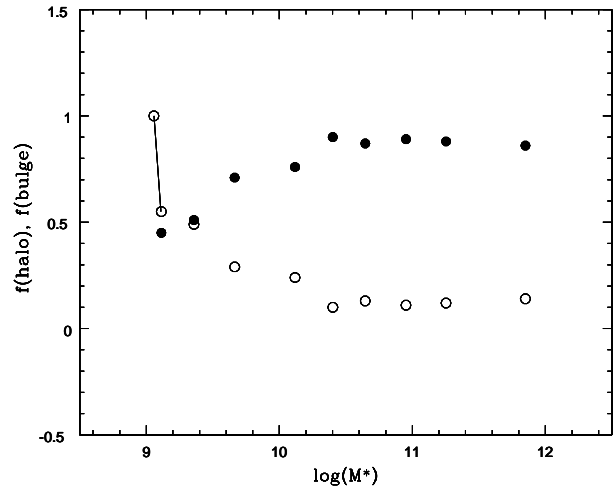}
\caption{Fraction of low-metallicity halo (open dots) and bulge-like stellar 
population (solid dots) as a function of total stellar mass of the fiducial 
galaxies. Red globulars were not detected in the lowest mass galaxy 
($log(M_{*})\approx 9.0$). Note that the fractions of halo and bulge stellar 
masses are almost equal at $log(M_{*}) \approx 9.3$.}
\label{fhalo_bulge_bis}
\end{figure}

The dependences of the mass of the low-metallicity stellar halo (shifted by +7;
open circles) and that of the bulge-like component (filled circles) with total 
stellar mass, as well as their respective mass fraction contribution, are
displayed in Figure\,\ref{mhalo_mbulge_mt} and Figure\,\ref{fhalo_bulge_bis}. 
The bulge-like component rises rapidly with galaxy mass, reaching 50 
percent at $log (M_{*}) \approx 9.3$ and about 85 percent in the three 
highest mass fiducials. In the low-mass galaxies, the $Z_{sr}$ parameters are 
low, i.e., the corresponding red GCs are in fact rather blue (although we identify 
them as ``red'' as a consequence of the working definition we adopted).


\begin{figure}
\includegraphics[width=\hsize]{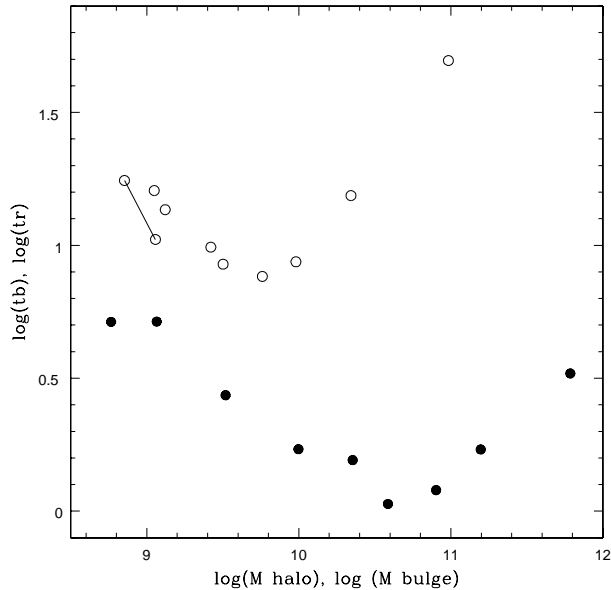}
\caption{Intrinsic globular cluster formation efficiencies $t_b$ (open 
circles) or $t_r$ (filled circles) as a function of the stellar masses of the 
halo or the bulge. Red globulars exhibit a marked minimum at $log(M*)=10.5$, 
coincident with the peak of the stellar surface density $\Sigma$. Both GC 
sub-populations reach minimum values of their $t$ parameters around the 
fiducial galaxy number 4.}
\label{t_intrin}
\end{figure}

The GC formation efficiency, $t$, was defined in terms of the total 
stellar mass. The next step imply the definition of an ``intrinsic'' $t$ 
parameter, i.e., the number of GCs per stellar unit mass of the associated 
(halo or bulge) component. These $t_b$ or $t_r$ parameters are depicted as a 
function of the halo or bulge stellar masses in Figure\,\ref{t_intrin}. Both 
GC sub-populations show a similarity in the sense that these parameters 
exhibit {\it U-shaped} forms and reach minimum values in the range defined
by the fiducial galaxies number 4 and 5.

\begin{figure}
\includegraphics[width=\hsize]{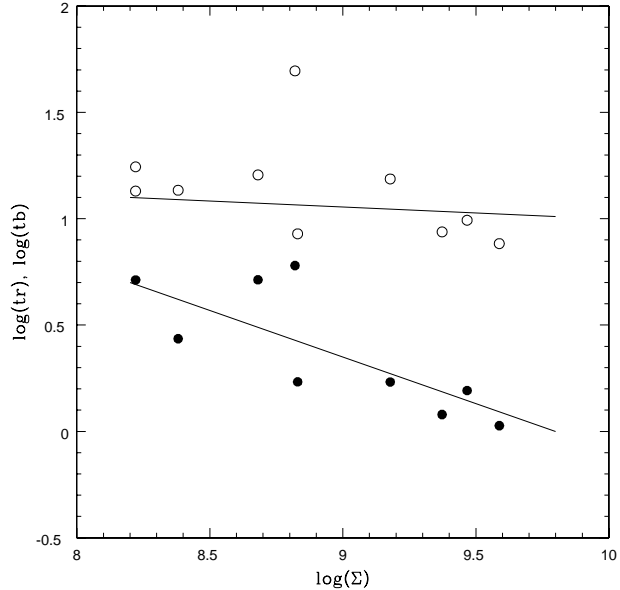}
\caption{Intrinsic globular cluster formation parameters $t_b$ for the blue 
(open dots) and $t_r$ for the red GCs (filled dots) as a function of the 
stellar surface mass density $\Sigma$. Both cluster families show a distinct 
behaviour. There is a clear trend shown by the red clusters whose $t_r$ 
parameters decrease with increasing $\Sigma$). In turn, blue globulars show a 
marginal variation and  a larger dispersion of their $t_b$ parameters. The 
highest value, $t_{b}=1.63$, corresponds to the fiducial galaxy with the
highest mass (number 1).}
\label{ttbbrr_surf}
\end{figure}

An intriguing difference is shown in Figure\,\ref{ttbbrr_surf}. In this 
figure, the red GCs (filled dots) show a clear anti-correlation with 
$log(\Sigma)$ which is practically absent for the blue GCs. This behaviour
is explained by the dominant role of the bulge-like component (connected with 
the red GCs) in terms of the mass of the projected stellar density $\Sigma$. 
In general, a dependence of the $t$ parameter with both galaxy mass and 
$\Sigma$ may be expected.  

\begin{figure}
\includegraphics[width=\hsize]{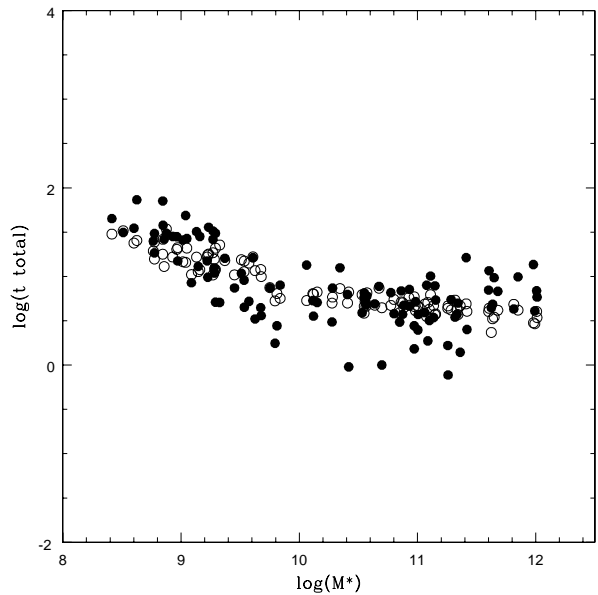}
\caption{Total globular cluster formation efficiency versus stellar mass for a 
sample of 126 E and S0 galaxies from \citet{HAR13} (filled dots). Open dots 
show the projection of these galaxies on the plane defined by the stellar mass 
and the projected stellar mass density $\Sigma$ (see text).}
\label{ttot_mt_Harris}
\end{figure}

As a first approximation, such a dependence is shown by a 2-D least-square fit 
to a sub-sample oh the HHA13 galaxy sample. Selecting galaxies with E or S0 
morphologies, within 30 Mpc of distance an more than 10 GCs yields:

\begin{eqnarray}
\label{eqn13}
log(t)=-0.29(\pm0.12)~log(\Sigma)-0.16(\pm0.07)~log(M_{*})+ \nonumber\\
6.02(\pm0.35)
\end{eqnarray}
                         
\noindent with a rms of $\pm~0.32$ and corresponding to 118 galaxies, 94 with 
genuine dynamical masses, and other 24 with masses derived through the 
$M_{dyn}-L_V$ calibration given by HHA13. For all the objects we assume 
that, half of the dynamical mass is contained within the effective radius of 
the galaxy, in  order to compute the corresponding $\Sigma$. A morphological
discrimination between E and S0 galaxies leads to practically equivalent fits.
The observed parameters ($t$ and $M_{*}$) and the values corresponding to the 
fit of a plane (open dots) are displayed in Figure\,\ref{ttot_mt_Harris}.


\section{Dark matter: with or without you?}
\label{LOWMETDARK}

\begin{figure}
\includegraphics[width=\hsize]{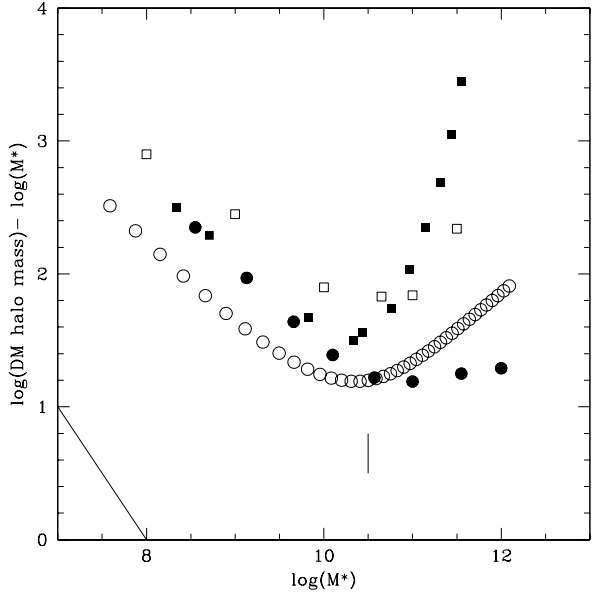}
\caption{Comparison of the dark mass to stellar mass ratio from different 
sources. {\it Filled dots:} \citet{SHA06}; {\it filled squares:} \citet{BER10};
{\it open squares:} \citet{LEA12} (see text); {\it open dots:} \citet{HAR13}. 
The line at left shows the effects of errors on stellar masses. The vertical 
line is an approximate position of a common minimum for all these ratios at 
$log(M*)=~10.5$.}
\label{dark_4}
\end{figure}

The connection between GCs and dark matter (DM) halos has been suggested in a 
number of papers in the literature (e.g. \citealt{BLA99}, \citealt{MCL99}, 
 \citealt{SPI08}, \citealt{GEO10}, \citealt{SPI09}, HHA13). A still 
problematic issue in this kind of analysis is the wide range of values 
concerning the ratio between dark halo mass and stellar mass.
Figure\,\ref{dark_4} compares different results and is illustrative of this 
situation. The data correspond to \citet{SHA06}, \citet{BER10}, \citet{LEA12} 
(values for redshifts lower than 0.1, compiled in their figure 10), and HHA13. 
In this last case, the authors adopt a parametric approximation that assumes a 
constant ratio of the total stellar mass in GCs to dark halo mass, 
$\eta=6 \times 10^{-5}$.

Despite the differences between these relations, they have two common 
features: on one side, the slopes are very similar for stellar masses below 
$log(M_{*})\approx10.5$. On the other, all of them exhibit a more or less 
evident minimum at this last mass. \citet{LEA12} identify this minimum as a
``pivot mass'' that changes with redshift but keeps a constant dark to stellar 
mass ratio $\approx27$.

In particular, and in a cautious way, we adopt the results from 
\citet{SHA06} for the following discussion. The galaxy stellar masses derived 
by these authors and ours follow a linear relation:

\begin{equation}
\label{eqn14}
log(M_{*})=0.768 + 0.937~log(M_{*~Shankar})
\end{equation}

\noindent and their galaxy masses (and dark mass to stellar mass ratios) were 
transformed to the galaxy mass scale of this paper in what follows.   

\begin{figure}
\includegraphics[width=\hsize]{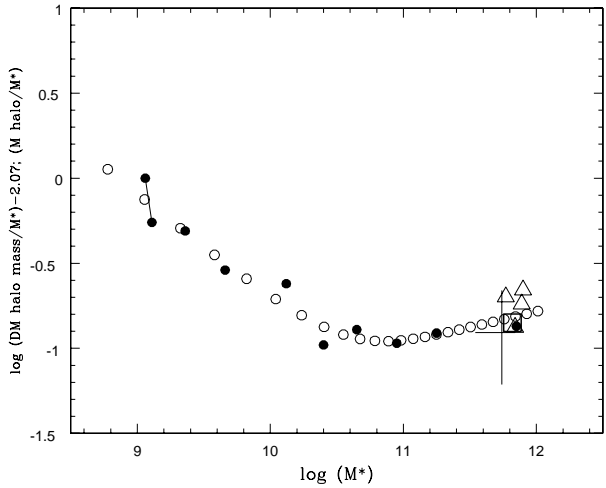}
\caption{Dark matter mass to total stellar mass ratio from \citet{SHA06} (open    
circles), values for NGC\,5846 from \citet{NAP14} (open triangles), and
for NGC\,4486 from \citet{AGN14} (cross). All of them shifted vertically by -2.07 in
ordinates. Low-metallicity halo mass to total stellar mass ratio for the
fiducial galaxies are shown as solid dots. The open square represents the
low-metallicity halo to total stellar mass in NGC\,4486 (based on FVF12).}

\label{halo_dm_mt}
\end{figure}

That choice is motivated by two arguments. First, 
Figure\,\ref{halo_dm_mt} shows a remarkable similarity between the dark to 
stellar mass ratios from these authors (scaled down by -2.07 in the 
logarithmic ordinate), and the ratio between the mass of the low-metallicity 
halos to total stellar mass from Table\,\ref{Table_4}.
Moreover, recent studies of the giant elliptical galaxy NGC\,5846, 
presented by \cite{NAP14}, and for NGC\,4486 by \citet{AGN14}, give a range 
of values of the dark halo to stellar mass ratios that are in excellent 
agreement with the \citet{SHA06} ratios.

Second, this is consistent with previous results based on the similarity of 
the spatial distributions of the blue GCs and low-metallicity halos and that 
of dark matter in NGC\,1399 and NGC\,4486 (FFG05 and FVF12, respectively).
The shift of the \citet{SHA06} relation in  Fig. \,\ref{halo_dm_mt}, 
corresponds to stellar masses of $\approx0.8\times10^{-2}$ of the dark halo 
mass. This is comparable to the ratio derived for NGC\,4486, by combining the 
mass of the low-metallicity halo given in FVF12 ($0.95\times10^{11}~M_{\odot}$), 
and the total enclosed mass within about 100 kpc from \citet{GEB09} and 
\citet{MUR11} ($1.3\times10^{13}~M_{\odot}$), that yield a ratio 
$\approx0.5\times10^{-2}$. 

Although these coincidences are encouraging for further analysis, we also note 
the discrepancy, pointed out by \citet{STR11}, between different estimates of 
the dark halo mass in NGC\,4486. Eventually, the adoption of the \citet{SHA06} 
relation, and our galaxy stellar masses, can be used to obtain the dark mass 
content for each fiducial galaxy, and for the definitions of $t_{dark}$, 
$t_{b~dark}$ and $t_{r~dark}$, i.e., the ratios between the number of total, 
blue and red GCs to dark matter mass.

\begin{figure}
\includegraphics[width=\hsize]{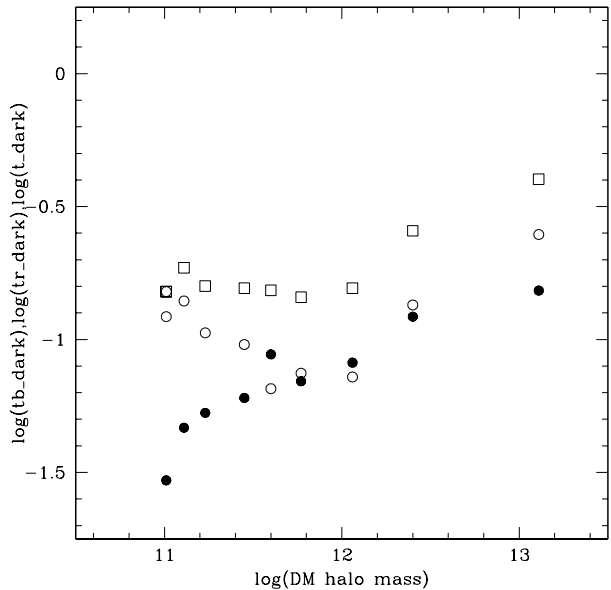}
\caption{Number of globular clusters per unit dark mass for the blue 
(open circles), red (filled circles) and total population of globular clusters 
per unit dark mass (open squares) as a function of dark halo mass.}
\label{all_t_dark}
\end{figure}

Figure\,\ref{all_t_dark} shows the behaviour of $t_{dark}$ as a function of 
the dark halo mass and the contribution of each GC sub-population to this 
quantity. Blue and red clusters show opposite trends in the dark mass range 
from $log(M_{dark})=11.0$ to 12.0 (i.e., galaxies fainter than the ``gap'' in 
the colour-magnitude diagram of Figure\,\ref{halo_dm_mt}), that once combined, 
lead to a rather constant $t_{dark}$ in that mass domain. 

\begin{figure}
\includegraphics[width=\hsize]{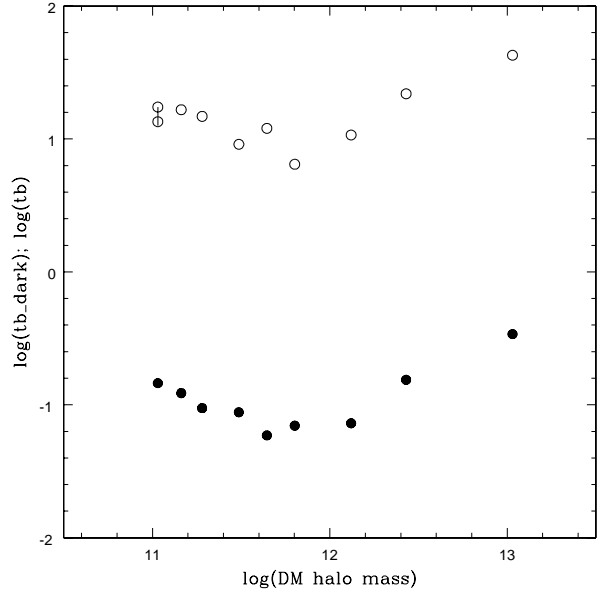}
\caption{Number of blue globular clusters per low-metallicity halo mass 
(open circles) and unit dark mass (filled circles) as a function of dark halo mass.}
\label{tb_tb_dark_dm}
\end{figure}

Figure\,\ref{tb_tb_dark_dm} compares the number of blue GCs per 
low-metallicity halo mass and per dark matter mass. These two parameters show 
similar trends (as expected if the low-metallicity halo and dark mass halo 
keep a constant ratio, as shown in Figure\,\ref{halo_dm_mt}). The minimum 
values in both $t_b$ and $t_{b\_dark}$ occur at the fourth fiducial galaxy, 
where the surface stellar density $\Sigma$ reaches a maximum.

\begin{figure}
\includegraphics[width=\hsize]{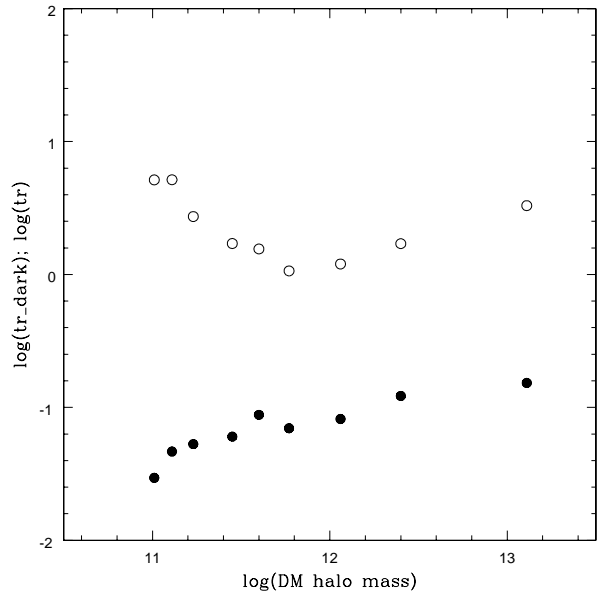}
\caption{Number of red globular clusters per bulge-mass (open circles) and 
unit dark mass (filled circles) as a function of dark halo mass.}
\label{tr_tr_dark_dm}
\end{figure}

This trend contrasts with the counterpart corresponding to the red GCs, shown 
in Figure\,\ref{tr_tr_dark_dm}. In this case, the number of GCs per dark mass 
units increases along with the mass of the dark halo, showing an inflection at 
$log(DM~halo~mass)\approx12.0$. Instead, the $t_r$ parameter shows a minimum 
at a value coincident with the mass of the galaxies where the surface stellar
density $\Sigma$ reaches a maximum.

\begin{figure}
\includegraphics[width=\hsize]{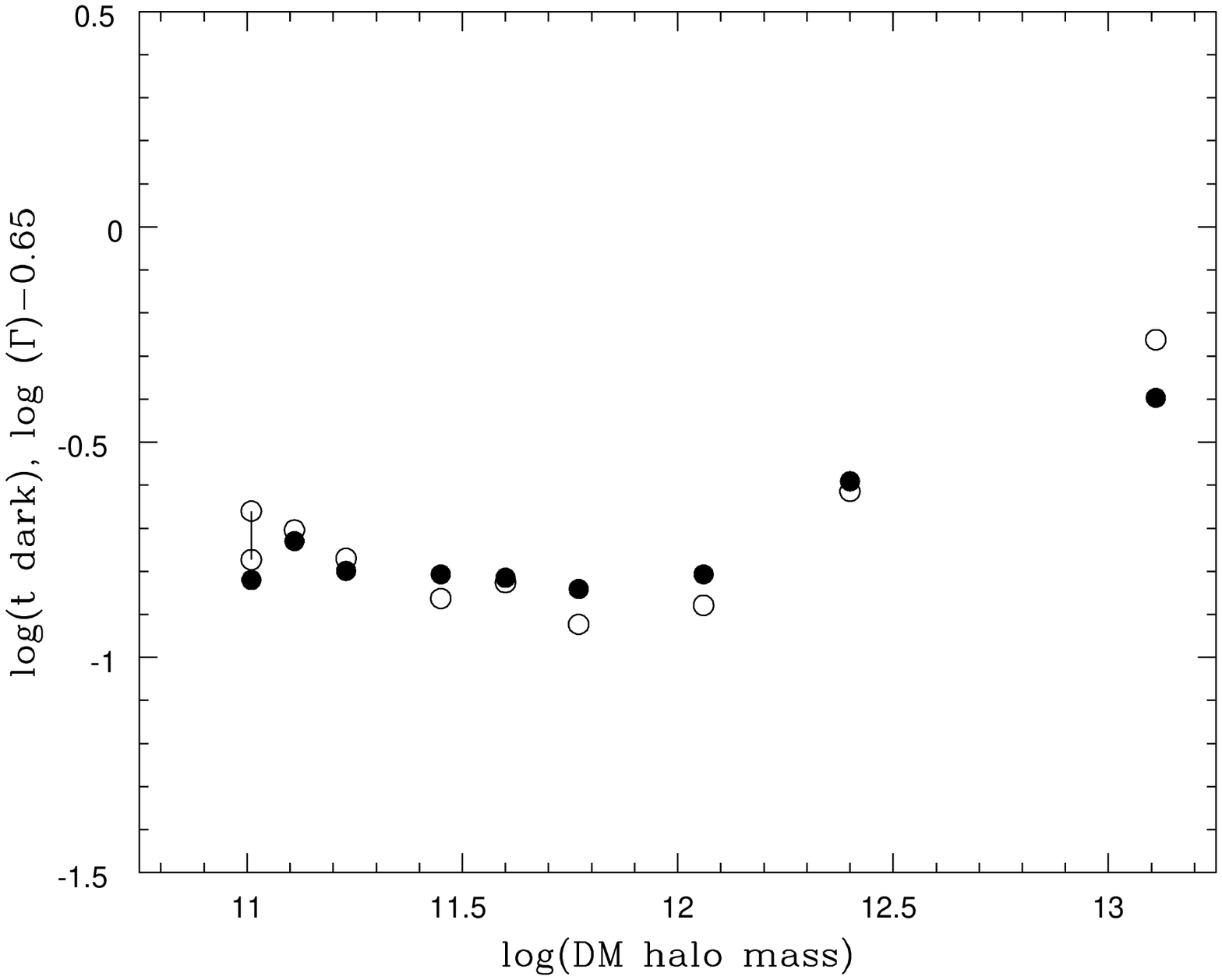}
\caption{$log(\Gamma)$, shifted by -0.65 (open circles), and $log(t_{dark})$ parameters 
(filled circles) as a function of dark halo mass for each fiducial galaxy.}
\label{gamma_tdark}
\end{figure}

For dark halo masses larger than $log(M_{dark})\approx12.2$, both $t_r$ and 
$t_{r~dark}$ rise in similar ways, although this is a consequence of our 
adopted approach regarding the distribution of blue and red GCs in the 
extended halos of the three most massive fiducial galaxies.
Figure\,\ref{gamma_tdark}, shows that the $\Gamma$ parameter listed in 
Table\,\ref{Table_4}, scales very well with $t_{dark}$ (through a
 shift of  $-0.65$ in the logarithmic ordinate). This connection suggests 
that, the proportionality coefficient $\Gamma$ may be an indication of the 
number of eventual GC formation sites (dark mini-halos) which will be occupied 
by blue or red GCs, depending on the availability of baryonic matter and its 
density (in turn associated with chemical abundance).

Finally, we note that \citet{WU13} have a completely different view of the 
situation in the frame of Milgromian mechanics. These authors argue that the 
$S_n$ parameter has two, and opposite, dependencies upon the apparent virial 
mass (including a ``phantom'', i.e., non-existent dark halo) that, combined, 
leads to the {\it U-shaped} relation with the stellar mass of the galaxy.


\section{Low mass galaxies: The case of the Fornax Spheroidal galaxy}
\label{Fornax}

The Fornax spheroidal galaxy is $\sim20$ times less massive than our lowest 
mass fiducial galaxy and has five GCs. \citet{BAT06} detect the presence of 
several stellar populations (ancient, intermediate and young) with a strongly 
assymetric metallicity distribution of the field stars, that peaks at 
$[Fe/H]\approx-1.0$, and has an extended tail that reaches 
$[Fe/H]\approx-2.5$. \citet{LAR12} emphasize that this distribution is clearly 
different, and seemingly could not be reconciled with that of the GCs. Four of 
these clusters have $[Fe/H]$ in the range from -2.1 to -2.5, while the 
remaining one (Fornax\,4), has a higher metallicity, $[Fe/H]=-1.4$. 

Establishing a connection of GCs with field stars is clearly difficult on the
basis of a few clusters and is certainly affected by stochastic effects. 
Besides, the star formation history of the Fornax dwarf seems rather complex 
(see, for example, \citealt{DEB12}). With these caveats in mind, we attempted 
the same approach used in this paper. We start with a blue GC population
characterized by a parameter $Z_{sb}=0.02$, similar to those in the lowest
mass fiducial galaxy, and change both the number ratio of blue to red GCs, as 
well as the  $Z_{sr}$ and $\delta$ parameters, aiming at reproducing the 
observed chemical abundance distribution of the field stars.

This is accomplished by adopting $Z_{sr}=0.13$, a number ratio of blue to red 
GCs of 4, and $\delta=4$, i.e., about twice the value that gives a proper fit 
to the more massive fiducial galaxies. We have no information about the 
behavior of the $\delta$ parameter for galaxies with masses below 
$log(M_{*})\approx9$ but an increase with decreasing galaxy mass, and stellar 
density, cannot be dismissed.
 
\begin{figure}
\includegraphics[width=\hsize]{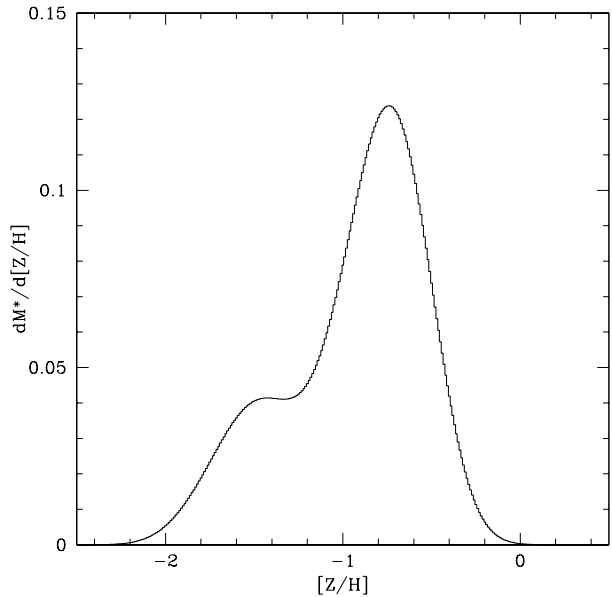}
\caption{Model chemical abundace mass spectrum for the stellar population in a
Fornax-like dwarf galaxy convolved with a gaussian kernel with a dispersion of 
0.2 dex in $[Z/H]$.}
\label{fornax}
\end{figure}

The  chemical abundance-stellar mass spectrum delivered by the model is 
depicted in Figure \ref{fornax} and shows a very good qualitative agreement
 with the  distribution for stars displayed in figure\,1 of \citet{LAR12}. 
This comparison assumes that the stellar mass contained in each chemical 
abundance bin scales directly with the number of stars in the same bin. 

Note that in the case of the Fornax galaxy we expect the presence of a single 
red GC, that possibly did not form or that eventually could be identified with 
Fornax\,4.
 

\section{Conclusions}
\label{CONCLU}

This paper revisits the connection between GCs and the stellar populations of 
67 galaxies included in the ACS Virgo Cluster Survey \citep{COT04}. The main 
results are:

\renewcommand{\labelenumi}{\arabic{enumi})} 
\begin{enumerate}

\item  The definition of composite {\it fiducial} galaxies was useful in 
clarifying the quantitative connection bteween GCs and the stellar populations 
of the galaxies they belong to. These fiducial galaxies are just a 
representation of average properties that arise as a result of different 
``nature-nurture'' events that take  place along the life of a galaxy. The 
possibility of reproducing the integrated colours of galaxies through the 
analysis of their GCs, and with a constant parameter $\delta$ 
(Equation\,\ref{eqn5}), strongly suggests the existence of a common mechanism 
that regulates the link between clusters and field stars. The value of that 
parameter indicates higher GC formation efficiency as metallicity decreases. 
 This dependence may be indicating a background process connected with 
enviromental conditions (through a metallicity-stellar density link) as noted 
in Section\,\ref{HALOBULGE}. 

We stress that the proposed GCs-field stars link also gives a proper 
representation of the chemical distribution of field stars for low-mass
galaxies (e.g. the Fornax dwarf spheroidal) although it requires higher 
$\delta$ values, something we cannot justify yet on observational basis. \\
 
\item The separation of GCs in two families, according to their respective 
chemical scale lengths $Z_s$, shows that blue and red clusters increase their 
number with total stellar galaxy mass but with distinct behaviours. Once 
combined as a ``single population'', the $log(N_{GCs})$ versus $log (M_{*})$ 
relation found for the Virgo ACS galaxies, is consistent with the non-linear 
trend (in a {\it log-log} plane) pointed out by HHA13. \\

\item The mean total stellar mass of the fiducial galaxies follows a linear 
relation with the {\it dynamical mass}, given in terms of the stellar 
dispersion and effective radius presented by \citet{WOL10} and correlated 
with $L_V$ in HHA13. \\
  
\item The empirical approach proposed to connect GCs and field stars allows 
an estimate of the low-metallicity halo and bulge-like stellar masses. In 
particular, these last structures arise in galaxies with stellar masses larger 
than $\approx10^9~M_{\odot}$ and grow up rapidly for more massive galaxies.

In this sense, they appear to be more ``pseudo-bulges'' (originated in the 
ability of the galaxies to retain their chemical outputs) than bulges that 
grow as a consequence of environmental effects.  

 We also emphasize that bi-modality is a very solid feature since 
``dry merging'' of any of the bimodal fiducial galaxies ends up as a new, but 
also bimodal, GCs colour distribution, although with different $Z_{sb}$ and 
$Z_{sr}$ parameters. \\
            
\item The behaviour of the number of the blue or red GCs per stellar mass 
units of the populations they are respectively associated (halo or bulge-like) 
as a function of total stellar mass, exhibits a similarity and a difference. 
On one side, they are both {\it U-shaped} and reach minimum values in the mass 
range defined by the fiducial galaxies 4 and 5. On the other, the red GCs show 
a marked (inverse) dependence with surface stellar mass density. This 
dependence is not detectable for the blue GCs as the total stellar mass is 
dominated by bulge like stars. \\ 

\item The ratio of the low-metallicity stellar halo mass to total stellar mass 
exhibits a strong similarity (once properly normalized) with the ratio of 
total dark matter mass to stellar mass as presented in \citet{SHA06}. In some 
way, the low-metallicity halo seems a kind of ``subtle echo'' of the dark matter 
content if the dark to stellar mass ratios given by \citet{SHA06} are adopted. 
            
\end{enumerate}

It is intriguing that the increase of the $\Gamma$ and $t_{dark}$ trends with 
galaxy mass, occurs approximately at the so called {\it critical mass} as 
defined by \citet{DEK03}; the mass value that suggests a ``dichotomy'' in the 
galaxy formation process as argued by \citet{KOR09}; the ``pivot'' mass defined 
by \citet{LEA12}; and, as pointed out in this paper, the stellar mass where 
the projected stellar surface density $\Sigma$ reaches a maximum value. 
               
The reason for that increase is not clear although the dramatic change of the 
effective radii of these galaxies with brightness (and stellar mass), leading 
to low stellar densities, suggests that this kind of environments may favour 
the GCs formation efficiency (speculatively, through better survavility 
conditions for the cooling flows where GCs would form).  

In fact, the hierarchical models by \citet{OSE10} indicate that the most 
massive galaxies, exhibit less in-situ star formation, have less concentrated 
halos, and also lower central densities. These authors also show that the size 
of these galaxies, and a large fraction of their masses, are the result of the 
accretion of objects formed ``far'' from the central regions of the galaxies. 

In our approach, the $Z_s$ scale lengths are a measurement of the spread 
of the metallicity of the blue GC/low-metallicity halo and of the red GC/bulge 
stars. The distinction between these ``two'' components is in fact a 
``working definition'' in the sense that they provide a good match to the 
integrated galaxy colours. Even though the low-metallicity component is rather 
homogeneous and would be properly identified as a ``single population'', this is 
not the case of the bulge-like component which, according to their larger 
$Z_s$ scales, suggests a rather inhomogeneous mix of stars with very different 
chemical abundances.  

 \citet{FORB97} already suggested that GCs bimodality may be the result of 
a ``two phases'' process. The meaning of ``phase'', however, seems still an open 
issue in the context of some recent results. For example, several works 
point out that re-ionization has played a role (e.g. \citealt{KAT12}, 
\citealt{SPI12}; \citealt{ELM12}; \citealt{GRI13}) in the formation of the  
metal poor blue GCs that  would appear at high redshifts and before the red 
(and more heterogeneous chemically) GCs.  \citet{TON13} (and also see 
\citealt{MUR10}) gives arguments to support the existence of a discrete 
temporal sequence in the frame of hierarchical models, indicating that blue 
GCs form at a redshift of $\approx4$ and are later accreted by galaxies that 
already have their own metal rich GCs (formed in-situ) at redshifts of 
$\approx2$. This is  similar to the scenario presented by \citet{FOR82} for 
the particular case of NGC\,4486 and later generalized by \citet{COT98}. 
\citet{TON13} also concludes that red GCs only form in galaxies with stellar 
masses larger than $10^9~M_{\odot}$, in approximate agreement with the results 
presented in this paper regarding to these clusters and their associated 
bulge-like structures.

In contrast with that landscape, and as noted by \citet{LEA13} (and also see 
\citealt{VAN13}) the results for the Milky Way GCs, rather than discrete 
events, shows a continuous and bifurcated age sequence. These results
indicate that both metal poor and metal rich GCs seem coeval and display the 
same range of ages ($\approx2$ Gy), although shifted in metallicity by  
$\approx0.6$ dex. In their view, blue GCs may need not to have formed in a 
truncated epoch and as separated episodes. A 2 Gy difference between the low 
(NGC\,6397) and the high metallicity (47\,Tuc) Milky Way GCs, is also
found by \citet{HAN13} on the basis of the study of the white dwarf sequences 
in these clusters.
 
In our analysis, we find that the chemical scale lengths of the blue GCs, 
$Z_b$, slowly increase with total galaxy mass, i.e., at least a fraction of 
them have a link with these galaxies. In a tentative explanation, and 
coincident with \cite{CARR13}, we suggest that blue GCs may be a mix of 
clusters formed in-situ (and showing a dependence with galaxy mass) and 
other accreted, more stochastically, from less massive fragments/galaxies.

If a bifurcated age sequence as in the Milky Way exists, the chemical 
abundances of both the blue and red GCs are genuine ``in phase'' clocks. This 
would indicate that the term ``phase'' (in the context of \citealt{FORB97}) is 
more a feature connected with environment than with time.
 
To the question posed by HHA13 regarding {\it ``what determines the size of a GC 
population?''}, this paper suggests that it is a competing mechanism where the 
number of GCs depends on the availability of potential formation sites, that 
increases with galaxy mass and, at the same time, has an inverse dependence 
with stellar density.  
     
As a final conclusion, we find support for the idea that GCs in fact codify 
information about the large scale features of ETGs. However, the complete code 
is not yet conclusively understood.

\section*{Acknowledgements} 
 We thank Dr. William E. Harris for  his interesting and helpful comments.
 JCF acknowledges the hospitality of Lic. Luc\'ia Send\'on (Director) of the ``Galileo Galilei'' Planetarium (Buenos Aires). 
 This work was supported by grants from  CONICET (PIP-2009-0712) and from La 
Plata National University (G128), Argentina. AVSC acknowledges finantial
 support from Agencia
 de Promoci\'on Cient\'ifica y Tecnol\'ogica of Argentina (BID AR PICT 2010-0410).

\label{lastpage}

\end{document}